# A review on scanning photocurrent microscopy and its application to one- and two-dimensional materials

T. Serkan Kasırga[1,2] *

[1] National Nanotechnology Research Center – UNAM, Ankara 06800, Türkiye

[2] Department of Physics, Middle East Technical University, Ankara 06800, Türkiye

*e-mail: kasirga@unam.bilkent.edu.tr

**Abstract**

The electrical response of a material when illuminated with light is a key to many optoelectronic device applications. This so-called photoresponse typically has a non-uniform spatial distribution through the active device area, and the ability to spatially resolve the photoresponse enables an in-depth understanding of the underlying physical mechanisms. Scanning photocurrent microscopy (SPCM) is a method that allows the spatial mapping of the photoresponse by raster scanning a focused laser beam over the sample. SPCM is becoming more popular due to its simplicity and power in unraveling fundamental optoelectronic processes. In this review, first, we provide the fundamentals of SPCM to lay the basics for the subsequent discussions. Then, we focus on the literature that employs SPCM to identify the photoresponse of one- and two-dimensional materials. We discuss SPCM measurement results of common materials in detail and introduce a systematic approach to interpreting the SPCM measurements. We have given particular emphasis on the photothermal mechanisms that are excited by the focused laser beam and critically reviewed studies in the literature from the perspective of laser-induced heating of the electronic and the lattice degrees of freedom. Finally, we discuss the shortcomings of SPCM in determining the mechanisms leading to the photoresponse.

**Main Text**

## 1. Introduction

The light-induced electrical current generation in solid-state devices is an active research field with many potential technological prospects in energy harvesting and conversion. Despite our deep understanding of the photocurrent generation mechanisms in solids, device performance and efficiencies can be further improved by studying the interplay among different mechanisms. In particular, low-dimensional materials offer unique avenues to explore light-matter interactions and photocurrent generation mechanisms. Extremely confined electromagnetic fields of plasmonic, excitonic, phononic, or magnonic origin in low-dimensional materials can be coupled with light and result in novel physical mechanisms to dominate the photoresponse. As a result, unconventional photoresponse mechanisms may be observed and even dominate the total photoresponse in low-dimensional materials, which may be harnessed to increase the overall quantum efficiency of the devices.

The position and distribution of the photoresponse generated on a device can hint at the underlying physical mechanisms. Moreover, the photoresponse can be modulated to uncover further information about the mechanisms by tweaking degrees of freedom in the incident light, such as the wavelength, polarization, and intensity. Scanning photo-current/voltage microscopy (SPCM) can provide a suitable experimental platform to achieve the abovementioned objectives. Using SPCM, photoresponse and optical intensity maps can be created simultaneously, and the location and the sign of the photoresponse can be resolved at the diffraction limit. The laser-beam induced current (LBIC) concept has been well-known since the early days of lasers; however, in the context of SPCM, it was first introduced in 1978 by Lang and Henry at the Bell Laboratories [1] to investigate the spatial distribution of the recombination centers in semiconductors. Over the past four decades, many research groups have contributed to the improvement of the method. Also, the advent of fast computers, data acquisition techniques, and lock-in amplification schemes has tremendously improved the sensitivity and implementation of scanning photocurrent microscopy.

This review focuses on the use of SPCM on one and two-dimensional materials and discusses the interpretation of the studies in the literature. First, the basics of SPCM are introduced in sub-section 1.1. Then, a brief overview of how photoresponse emerges in semiconducting and metallic materials is given in the sub-section 1.2 to provide a fundamental basis for the subsequent discussions. Sections 2 and 3 provide an in-depth discussion of how SPCM has been used in determining the photocurrent generation mechanisms in some exemplary 1D and 2D materials, respectively, as well as the limitations of the SPCM by referencing the most relevant literature. A systematic approach to interpreting the SPCM measurements, limitations of SPCM, and how SPCM can be used in conjunction with other techniques to develop a better understanding of the photoresponse in low-dimensional materials are discussed in Sections 4 and 0, respectively. Covering SPCM studies both on 1D and 2D materials made the review very bulky. This led some of the studies that are not focused directly on the SPCM of 1D or 2D materials to be left out. Yet, we consider that discussing 1D and 2D under the same review is essential for constructing a general framework for understanding and interpreting the SPCM results in low-dimensional materials. As a result, at the cost of having a lengthy review, we discussed the most notable SPCM work on relevant materials in depth. As a final comment, we would like to emphasize that rather than simply summarizing the SPCM-related work, we provided a critical view of the results that were particularly relevant to the interpretation of the SPCM results. This critical view forms the basis of Section 4 where we discuss how to interpret SPCM results in a systematic way.

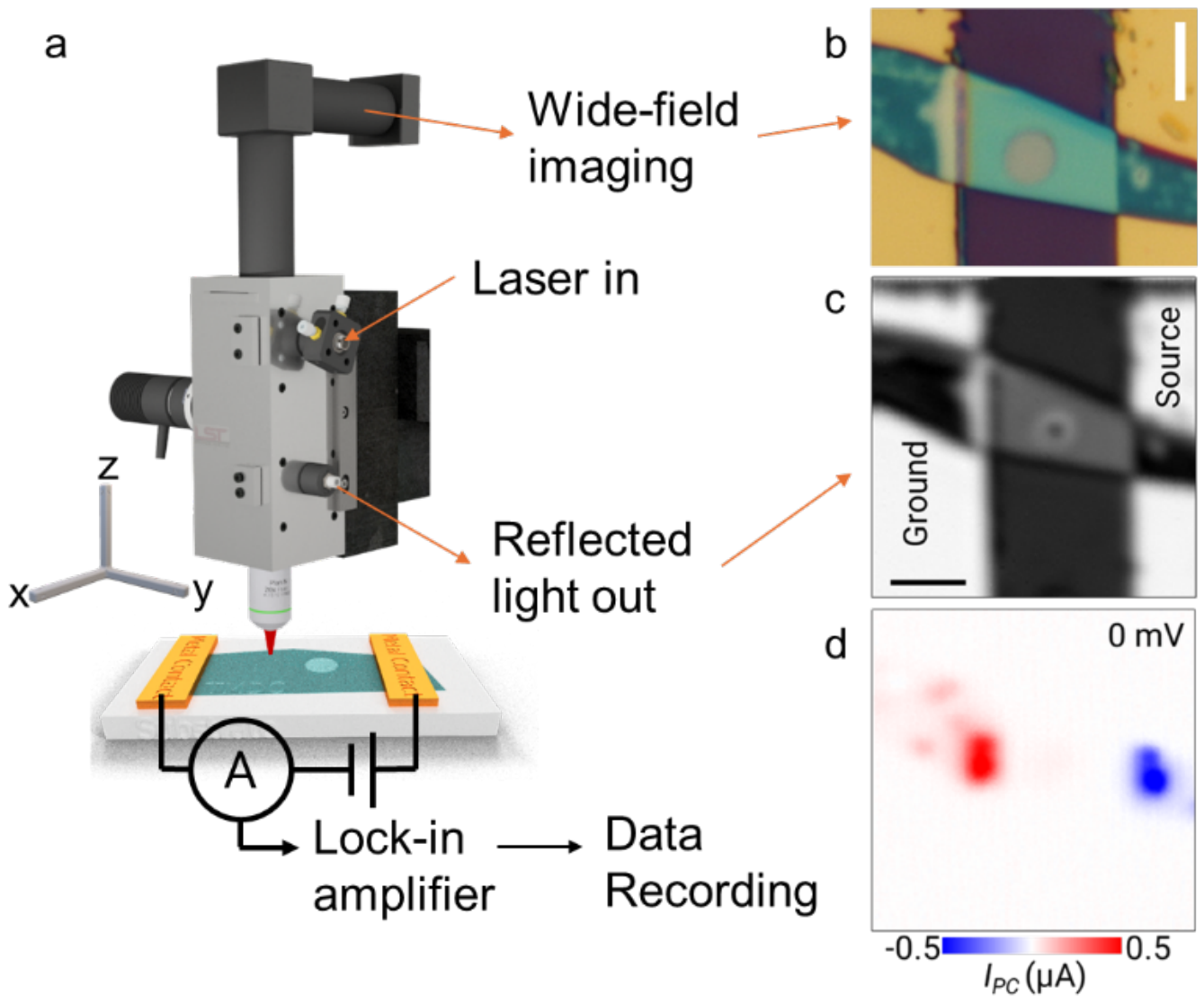

**Figure 1 a.** Schematic of scanning photocurrent microscopy setup depicts the essential components of SPCM measurements. **b.** An optical microscope image of a sample is shown. Scale bar is 5 μm. **c.** A greyscale reflection intensity map generated by SPCM is shown. Electrodes are labelled. Scale bar is 5 μm. **d.** The corresponding photocurrent map shows the spatial distribution of the light-induced electrical response. The sign of the signal depicts the direction of the photocurrent with respect to the current preamplifier.

### 1.1. Basics of SPCM

The essence of scanning photocurrent microscopy is the raster scanning of a focused laser spot over an electrically contacted material to collect light-induced current (or emf). Various raster scanning strategies are discussed later. To spatially correlate the photoresponse map with the device, the reflected light is collected by a photodetector. As a result, a reflected (or transmitted, in rare cases) light intensity map is created simultaneously. In a typical SPCM setup, lock-in amplification is used to isolate any DC contribution to the photocurrent. Moreover, a current pre-amplifier is often employed to pre-filter the signal as well as to eliminate impedance mismatch. Details regarding the electrical measurements and beam steering in an SPCM setup are discussed in **Appendix A**. A schematic of a typical SPCM setup and typical SPCM-generated maps are given in **Figure 1**. In the following subsections, the spatial and temporal resolutions of a typical SPCM setup and their impact on the measurements are discussed in detail.

#### 1.1.1. Spatial resolution of SPCM

For conventional optical systems, Abbe's diffraction limit sets the spatial resolution. The minimum resolvable distance can be expressed as $d = \lambda/2\mathrm{NA}$, where $\lambda$ is the light wavelength and NA is the numerical aperture. To achieve resolution at the diffraction limit, high-quality, high-magnification objectives (typically 40 to 100x) with large numerical apertures are used in conventional SPCM setups. To achieve a laser spot focused down to the diffraction limit, the laser beam must be shaped appropriately. For the free-beam optics, beam shaping can be achieved via a series of variable irises and beam collimators. For fiber-coupled lasers, single-mode fiber optics needed to be used to ensure smaller $M^2$ values. In multi-mode fibers, it is difficult to ensure uniform illumination of the input end of the fiber, which inevitably leads to much poorer $M^2$ values as compared to single-mode fibers. Also, to achieve higher spatial resolution, a Gaussian intensity profile is desired. Although there are methods to achieve resolutions beyond the diffraction limit, SPCM with these so-called super-resolution techniques is beyond the scope of this review [2–9].

SPCM records two essential data as a function of position: (1) the current/voltage, (2) the reflected light intensity. Both data are recorded in pixelated maps, where each pixel corresponds to a physical position on the sample. To be able to utilize the diffraction-limited resolution, scanning must be performed at the same or better step size between each pixel. Consequently, spatial variations within ~250 nm can be correlated with local material and device properties. The actual resolution of an SPCM setup can be determined from the reflection intensity map. The first derivative of the intensity profile with respect to position across a high-aspect-ratio, high-contrast feature with an edge sharper than the diffraction limit, such as the edge of a metal electrode, can be fitted to a Gaussian function. The width of the fitted Gaussian function gives the spot size, and half of it is the resolution of the measurement. ***Figure 2*** shows the FWHM of a laser spot determined on a calibration sample with the method described above. AFM is used as a check to confirm the steepness of the sample walls. Although AFM could do a much finer scan, we performed a 20 μm scan with 256 data points, which gives 78 nm resolution or 156 nm measurement width. It is important to have a measure of the spatial resolution for the analysis of the results, as interpretation of the results often requires precise knowledge of the spatial resolution.

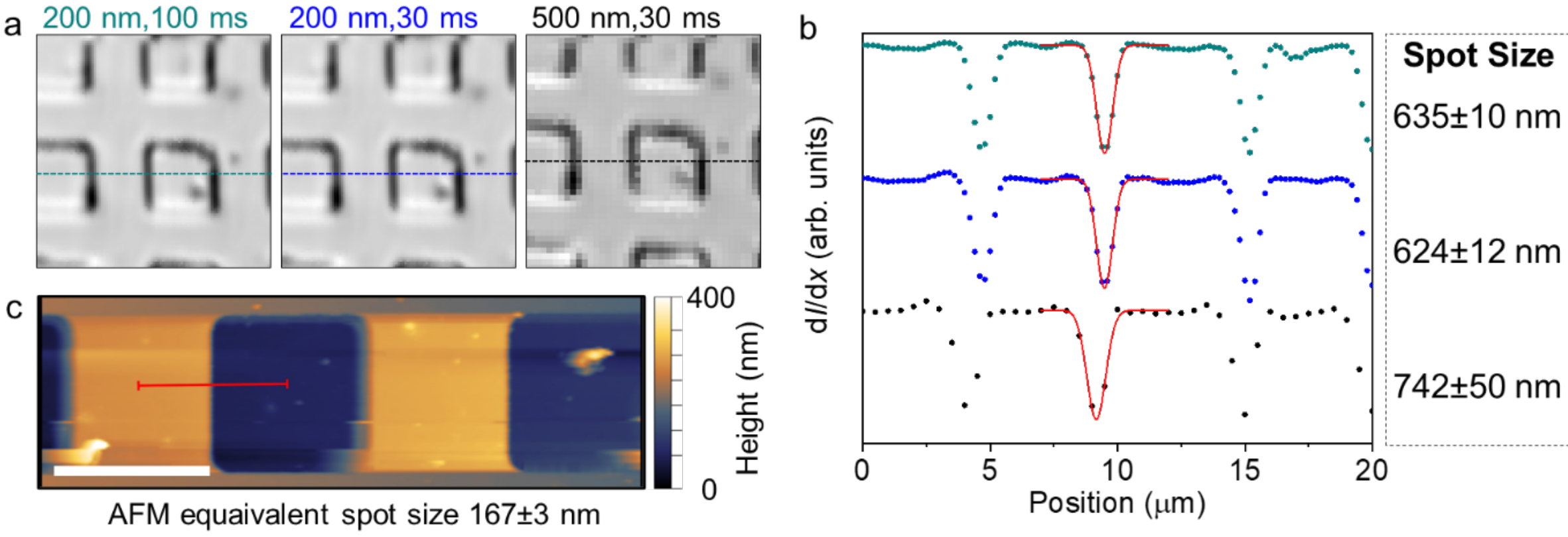

**Figure 2 a.** SPCM reflected light intensity maps collected at three different parameters on a calibration sample with 5 μm wide square grooves with 5 μm pitch. **b.** First derivative of reflected intensity with respect to position along the line traces shown in **a**. The width of the Gaussian function fitted to the peaks gives the spot size (or half of it gives the resolution) of the measurement. **c.** AFM height trace map of a 20 μm wide region of the sample with 256 points per line has a calculated equivalent spot size of 156 nm, and the measurement results give 167 nm, which shows that the sample walls are steep, but the loss of resolution is due to the diffraction limit in the SPCM.

#### 1.1.2. Temporal resolution of SPCM

The temporal resolution of SPCM reported in the literature is typically limited to a few milliseconds. In principle, there is no fundamental limitation to achieving femtosecond-short temporal resolutions. Each pixel can be measured in a time-resolved manner; however, achieving such an experimental setup requires fast lasers and fast lock-in amplifiers. Moreover, signal extraction will be limited due to the decreased LIA time constant in such a measurement. One way around the limitations of lock-in amplifiers in ultra-fast SPCM measurements is introduced by Sun et al. [10]. They used an ultra-fast laser with a 250-fs pulse width to pump the material. A probe beam is separated by a delay stage and chopped by a mechanical chopper at "regular" SPCM frequencies (i.e., ~kHz). The lock-in measures the chopped probe beam. When the pump and probe beams are temporarily separated enough from each other, the effect of the pump relaxes, and only the effect of the probe is measured. However, as the pulses are close enough, the pump has already excited enough electrons, and the signal from the probe saturates [10]. As a result, it has been possible to temporarily resolve the SPCM signal down to the laser pulse width. Recently, Silva et al. [11] reported time-resolved SPCM scans on topological insulator samples. Their method allowed fixing the pump beam to a point on the sample and scanning the probe beam. As a result, the probe beam could map out both spatially and temporally the effect of the pump over the sample. Despite the benefits, due to the complexity introduced by the pump-probe scheme, most SPCM on low-dimensional materials is performed within the "steady state", where time scales are much longer than the electronic and lattice thermalization timescales.

One aspect that requires particular attention when performing SPCM with LIA is that at each pixel, the laser beam needs to stop at the measured pixel for a brief period of time, at least as long as the time constant of the LIA, to prevent loss of spatial resolution. Spatial resolution is lost when the laser moves to the adjacent pixel, as the LIA is still performing the previous time integration, and multiple pixels can read the same average value, resulting in a loss of spatial resolution. ***Figure 2*a** demonstrates the spatial resolution loss, where the two scans have the exact same optical and scanning parameters but were recorded at different integration time constants.

### 1.2. Overview of photoresponse mechanisms in solids

Multiple physical mechanisms can contribute to the light-induced electrical current or emf generation in a material. The microscopic origin of the photoresponse mechanisms in metallic and semiconducting solids is different and typically happens at different time scales. Due to this difference, we investigated the photoresponse of metals and semiconductors under different subsections. Here, we will provide an overview of these mechanisms and how they may manifest themselves in an SPCM measurement.

#### 1.2.1. Photoresponse of metals

In metals, when a nondestructive ultra-short laser pulse hits a few tens of nm thin metal surface, some of the incident energy is absorbed by the interband transitions, and some is absorbed by the intraband transitions. Within a hundred fs, electron-electron scattering reconfigures the non-equilibrium energy distribution and creates a pool of "hot" electrons. At this stage, the electron temperature is far larger than the lattice temperature. On the order of a few picoseconds, the electron-phonon scattering equilibrates the two temperatures [12–16]. However, these timescales are typically very short for the SPCM measurements, as mentioned earlier. The hot electrons thermalize long before reaching the electrical contacts, as the typical device dimensions are much larger than the diffraction-limited spot size. The net effect of the laser illumination is then the heating of the material. For bulk metals, due to large thermal conductivities and multiple heat transfer pathways, the heat is dissipated efficiently through the degrees of freedom of the material before inducing any measurable electrical change [17].

Yet, the laser-induced temperature increase may cause a photothermal response in nanoscale metals, as the heat transport mechanisms and pathways can be modified dramatically as compared to the bulk case. Also, confinement in low-dimensional materials can inhibit mechanisms that cause the relaxation of hot electrons, as in the case of graphene, which can lead to unconventional photoresponse mechanisms [10,18,19]. Moreover, plasmonic effects can enhance the local heating even further in nanostructured metals [20–22]. Consequently, the local temperature increase can induce two current generation (or modification) mechanisms. The first mechanism occurs when thermoelectric junctions are formed between two metals (or regions of metals with dissimilar Seebeck coefficients). Photoinduced heating at such junctions can induce an electromotive force due to the Seebeck effect [23–26]. The second mechanism is the bolometric response, in which the temperature rise caused by the laser increases the material's resistivity; thus, when the metal is under bias, a negative photoresponse, i.e., a reduction in the electrical current, is observed [27]. The bolometric response is not a photocurrent generation mechanism, but rather a modification to the already flowing electrical current. These two mechanisms will be central to our discussion of the photoresponse observed in metals. Moreover, we will also briefly discuss how these two mechanisms can be used to measure the thermal conductivity of 1D or 2D metals in the following sections.

### 1.2.2. Photoresponse of semiconductors

In semiconductors, light-induced current generation is more prevalent. The separation of non-equilibrium charge carriers due to the built-in electric fields, namely the photovoltaic effect (PV), is a common mechanism in semiconductor device structures. Typically, the long minority-carrier diffusion lengths in semiconductors enable efficient photocurrent collection by electrical contacts at a distance. For instance, the room temperature minority carrier diffusion length in Si is on the order of tens of micrometers [28]. Consequently, micro-scale devices of Si can result in very large photocurrent generation. **Figure 3** depicts the contrast between an Ohmic and a Schottky junction between a 2D semiconductor and a 3D metal. In the case of a Schottky junction, effective photo-induced non-equilibrium charge carrier separation can be achieved, as shown for various source-drain bias and gate voltage configurations across the junction. Also, it is evident from these depictions that the contact type is important for the band alignment. All the relevant details regarding the PV effect and how it is related to the aforementioned parameters are discussed in detail in relevant 1D and 2D materials.

SPCM can provide a direct, spatially resolved method to evaluate the minority carrier diffusion length in semiconductors. The evaluation relies on analyzing the spatial decay of the laser-induced signal as the scanning laser spot moves away from the junction with the built-in electric field, such as a Schottky junction or a pn-junction. When electron-hole pairs are optically generated outside the depletion region, where the built-in electric field is negligible, minority carriers must diffuse toward the junction to be separated and collected. Assuming a one-dimensional diffusion model, the photocurrent decays exponentially with the distance from the junction interface $I_{PC}(x) \propto \exp\left(-\frac{|x-x_0|}{L_D}\right)$ where $x$ is the laser position and $x_0$ is the position of the junction, $L_D$ is the minority carrier diffusion length. By extracting lines perpendicular to the junction and fitting the decaying tail of the photocurrent peak to this exponential function, $L_D$ can be extracted. However, optical resolution limits and competing effects have to be taken into consideration when evaluating $L_D$. [29]

The photoconductive effect is another common mechanism in semiconductors where the applied bias drifts light-induced free carriers. Overall, due to this drift, the electrical resistance of the semiconductor is reduced, and larger electrical currents can flow under the same applied bias. The photoconductive effect may also emerge from the photo-charging of trap states within the material, as well as for 1&2D materials, photo-charging of trap states within the substrate. Electrons or holes populating localized trap states emerging due to defects, impurities, etc., can act as a local gate to modulate the conductance. Photogating and photoconductive effects can be distinguished by studying gate-dependent photocurrent measurements. Photocurrent from the photoconductive effect can be modulated instantaneously via an electric field. Whereas photocurrent from photogating will be temporally less responsive to the gating. Moreover, the difference in the response timescales of photoconductive and photogating effects might be used to distinguish the photocurrent generation mechanisms [30].

The photoresponse from a semiconductor can be enhanced by improving carrier collection efficiency via modified contacts. As we will discuss later, contact configuration can be important while assessing the SPCM maps. There are other mechanisms, such as the Seebeck effect, bolometric effect, and photogalvanic effect, that can lead to a sizeable photoresponse in semiconductor devices. In some cases, these effects can dominate the photoresponse. We will go into further details of these mechanisms in the relevant upcoming sections in detail.

Another important aspect we will emphasize in interpreting SPCM results for semiconducting 1D and 2D materials is that multiple photoresponse mechanisms can coexist and compete within the same device. It is exceedingly difficult to differentiate the mechanisms using SPCM only since any alterations in the material or the device, i.e., via application of electric field, strain, etc., perturb the other mechanisms as well. This mandates careful planning of controlled experimentation and analysis of the measurement results to elucidate the photoresponse mechanisms. For instance, in the case of semiconductors, the Seebeck effect is typically very large and modulable and may get intermixed with the PV signal. Signals generated by these mechanisms both appear near the contacts, and their spatial distribution has a similar line shape. Moreover, an external electric field modulates both PV and Seebeck effects [31]. In such cases, distinguishing the mechanisms requires a careful analysis of the SPCM measurements, and most often, a secondary method or a careful theoretical analysis of external electric field dependence is needed to differentiate the origin of the response. These challenges are discussed in detail in the relevant sections.

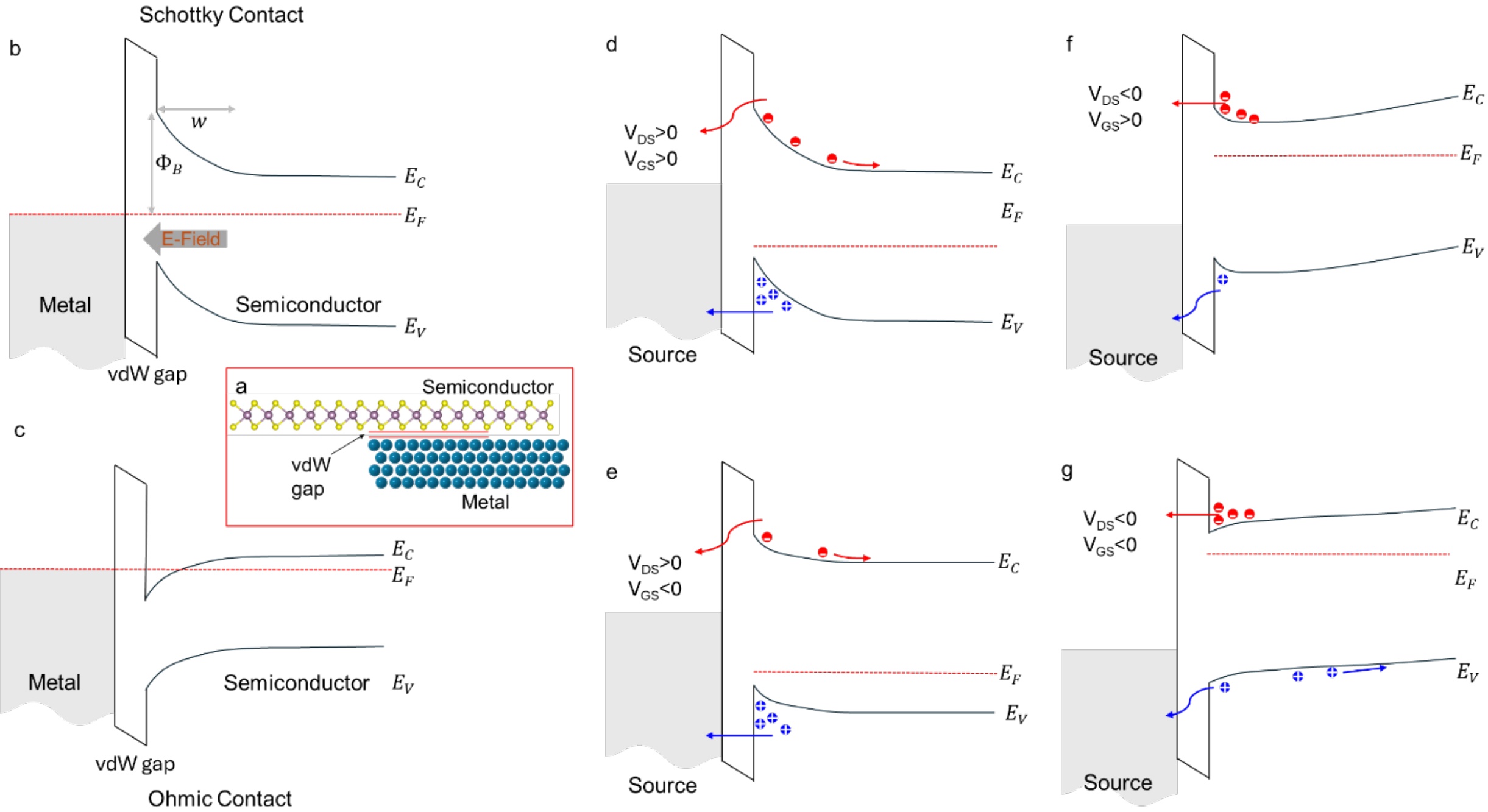

**Figure 3 a.** A junction between a 2D semiconductor and a 3D metal is depicted. The van der Waals (vdW) gap is marked between the materials. **b.** A Schottky-type junction and **c.** an Ohmic-type junction band diagrams are depicted. The depletion region width is denoted by $w$ and the Schottky barrier height is shown with $\Phi_B$. Built-in electric field is denoted with a solid arrow. **d-g.** How electrons (red) and holes behave in a Schottky contact under various source-drain bias and gate voltages. Depending on the configuration, the Fermi level and band alignments change.

In the following parts, we discuss the scanning photocurrent microscopy measurements reported in the literature on various 1- and 2-dimensional materials and devices. Based on these discussions, we will introduce a general framework to understand the application of SPCM on low-dimensional materials.

## 2. SPCM on one-dimensional materials

Size confinement on the electronic degrees of freedom results in the emergence of various mesoscopic phenomena [32]. Early attempts to create one-dimensional wires have been on 2D electron gas in high electron mobility transistor structures by creating confinement via local gating. Despite their exciting properties, these nanowires are typically not accessible by light as they are buried across various epitaxial layers; thus, they are beyond the scope of this review [33]. We will focus mostly on SPCM studies on carbon nanotubes and various semiconducting and metallic nanowires [34]. We will also discuss SPCM studies on nanometer-sized wires of strongly correlated materials where electron-electron and electron-phonon interactions dictate their electronic properties.

### 2.1. SPCM studies on Carbon Nanotubes

Following the synthesis of carbon nanotubes (CNTs) by Iijima [35] in 1991, there was a broad fascination with the amazing properties and tantalizing opportunities offered by these quasi-1-dimensional nanotubes. CNTs are rolled-up graphene sheets and can be found in semiconducting and metallic phases, depending on the sheet rolled to form the nanotube [36]. Also, the number of concentric tubes (single-walled or multi-walled) influences CNTs' properties. Carbon nanotubes (CNTs) can be classified into two types based on their electrical conductivity: metallic and semiconducting. Metallic CNTs exhibit metallic behavior with high electrical conductivity, while semiconducting CNTs show semiconducting behavior with an energy band gap and tunable electrical conductivity depending on their chirality and diameter. As a result, the photocurrent response of CNTs heavily depends on the type of CNT as well as the device configuration [37–39]. Over the past two decades, SPCM has played a crucial role in studying the origin of the photoresponse in carbon nanotubes [40–48].

Balasubramanian et al. [43] reported the first SPCM measurements on metallic single-walled CNTs and observed a bipolar photoresponse at the contact-CNT junction, i.e., photocurrent in the opposite direction at each contact (**Figure 4a-b**). 514.5 nm excitation laser at 1 mW (~150 kW/cm2) power is employed for the SPCM measurements. They concluded that the substantial photocurrent is due to the local energy barriers introduced when the CNT comes in contact with the metal electrodes. Despite their devices having Si as a back gate, they haven't performed any gate-dependent photoresponse measurements.

In 2007, both Freitag et al. [44], Ahn et al. [45] and Lee et al. [46] reported SPCM measurements on CNTs. Freitag et al. focused on the potential modulations along the carbon nanotubes due to the charging of the built-in defects. They haven't reported the electronic properties of the CNTs used in their experiments. Yet, their results are very similar to the first report by Balasubramanian [43], except for the photoresponse in the middle of the CNT (**Figure *4*c-e**). Ahn et al. [45] studied semiconducting CNTs and investigated the internal p-n junctions in ambipolar CNT transistors. Under zero gate voltage and zero bias, localized photocurrent spots near both electrodes appear with the opposite sign. They also performed gate-dependent measurements and illustrated that the sign of the photocurrent could be reversed depending on the gate voltage. SPCM on partially suspended CNT devices is also performed in the study. In this case, the CNT is partially suspended across the metal contacts, and strikingly, the photocurrent spots shifted to the ends of the suspended section. They attributed the observed shift to the modified dielectric constant of the environment that CNT experiences. Lee et al. reported very similar measurements with similar conclusions [46].

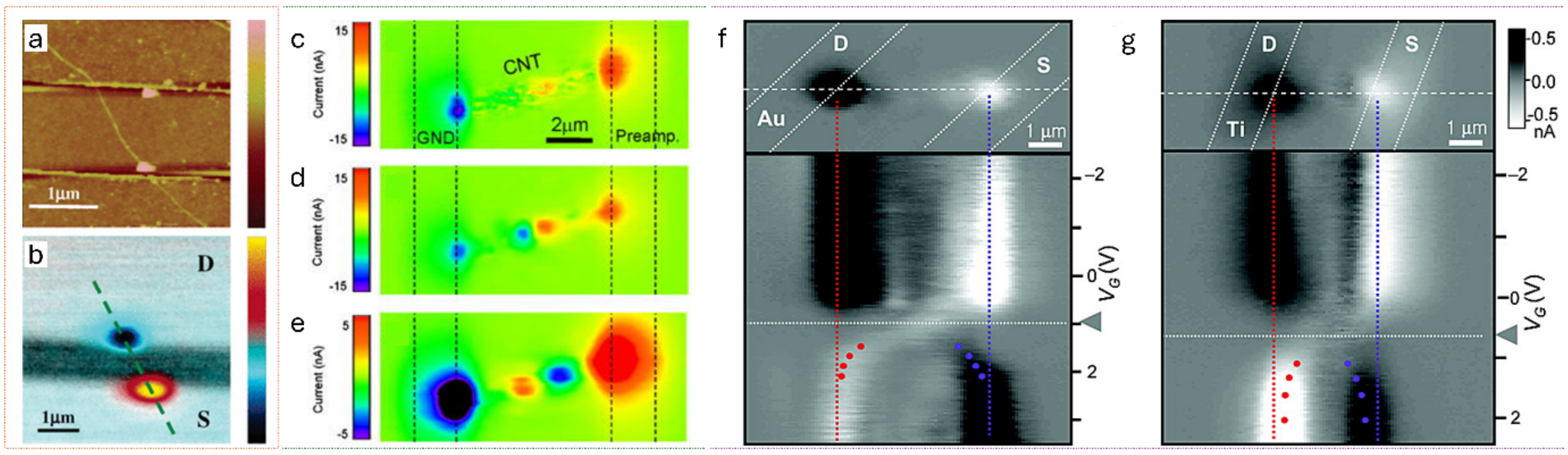

**Figure 4 a.** AFM height trace map of a sample showing a metallic CNT across two metal electrodes. The diameter of the CNT is 1.3 nm. **b.** SPCM photocurrent map overlayed on the reflection map is given. (S − source, D − drain, $\lambda_{exc}$ = 514.5 nm, 50× objective, 1 mW). Adapted with permission from Balasubramanian et al. [43]. Copyright 2005 American Chemical Society. **c.** SPCM maps of CNT without defect and **d.** with defect under zero gate voltage. **e.** SPCM map under -25 V gate voltage. The response at the center of the CNT is attributed to the defect. Adapted with permission from Freitag et al. [44]. Copyright 2007 AIP Publishing. **f.** and **g.** shows SPCM maps (upper panel) and SPCM line traces with gate voltage sweeps (lower panel) for two different semiconducting CNT devices. Adapted with permission from Ahn et al. [45]. Copyright 2007 American Chemical Society.

Gabor et al. [40] used SPCM and scanning photocurrent spectroscopy to show extremely efficient multiple electron-hole pair generation in split-gate field-effect CNT devices. By selectively tuning the gate voltages, they could produce p and n-doped regions, and zero-bias photocurrent can be observed. This work clearly demonstrated the role of photovoltaic response in a CNT field-effect device. In a follow-up study, Barkelid et al. [47] performed polarization-dependent photocurrent microscopy and spectroscopy on split-gate field-effect CNT devices (***Figure 5*****a-d**). They demonstrated the polarization dependence of the photoresponse and deduced the dielectric constant of the semiconducting CNTs based on their SPCM results.

In another study, Barkelid et al. [42] provided an in-depth analysis of the photocurrent generation mechanisms in suspended metallic and semiconducting CNTs in the split-gate field-effect configuration using SPCM. Their results unambiguously show that the photocurrent generated in metallic and semiconducting nanotubes has a photothermal and photovoltaic origin, respectively. For the metallic nanotubes, they demonstrated that the gate voltage could tune the Seebeck coefficient, and the response at the CNT-electrode interface changes signs depending on the gating. In a follow-up paper by DeBorde et al. [48], the photothermoelectric effect in suspended semiconducting CNTs was also demonstrated using SPCM. They showed that the thermal distribution along the CNT causes an emf generation due to the Seebeck effect as the temperature at the electrode metal and the CNT junction is larger than the other junction. In their measurements, they could distinguish the photothermal and photovoltaic responses (***Figure 5*****e-f**). Similarly, Buchs et al. [49] demonstrated that the dominant or non-dominant character of photovoltaic and photothermal mechanisms depends on the doping profile as well as the contact resistance.

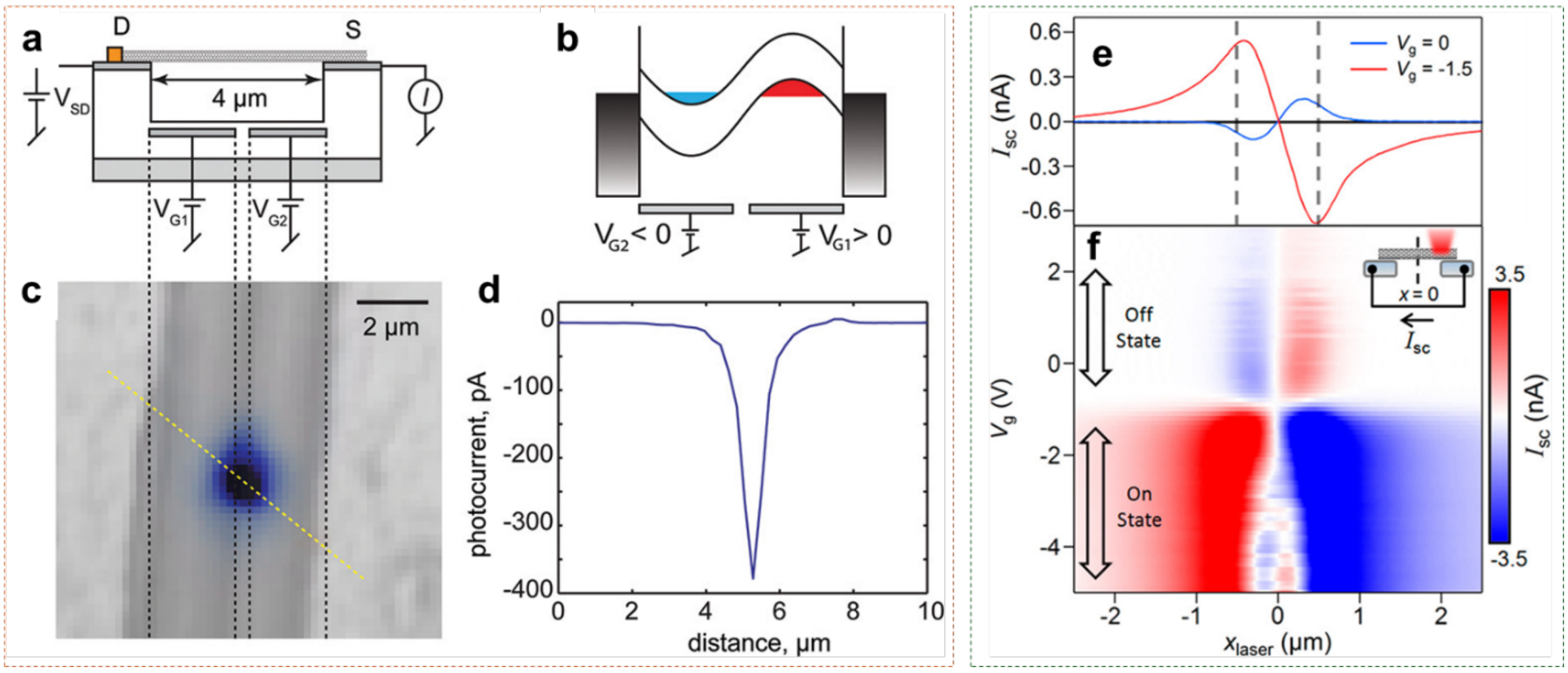

**Figure 5 a.** Schematic of a suspended carbon nanotube on two metal gate electrodes. **b.** Schematic of the spatial distribution of the electrostatic potential of the gate-defined p-n junction on a carbon nanotube. **c.** SPCM map overlayed on the reflection map, and **d.** is the line trace through the yellow dashed line. Reprinted with permission from Barkelid et al. [47]. Copyright 2012 American Chemical Society. **e.** Line scans of a CNT device under zero and -1.5 V global gate voltage. Vertical dashed lines indicate the position of the electrodes. **f.** Map of the gate voltage along the CNT. Adapted with permission from DeBorde et al. [48]. Copyright 2012 American Chemical Society.

In summary, CNTs exhibit various photoresponse pathways depending on their electrical properties. As opposed to the earlier studies, the most recent SPCM measurements on CNTs reveal that both photothermal and photovoltaic contributions are present in the photoresponse. These results illustrate that a systematic approach is required to understand the underlying photoresponse pathways. We would like to emphasize that, despite the small radius and optical absorption cross-section of the CNTs, relatively small laser power can induce a dominating photothermal response. This is informative to elucidate the often-ignored role of photothermal effects in the SPCM response of other low-dimensional materials.

### 2.2. SPCM studies on Silicon and other semiconducting nanowires

Silicon nanowires (Si NWs) have attracted a great deal of attention, perhaps more than CNTs, due to the exciting possibilities in optoelectronics and photovoltaics, the deep knowledge of Si processing, and its inherent compatibility with the CMOS processes. As a result, SPCM has been a widely used tool in studying the photoresponse of Si NWs. As anticipated, the SPCM-based studies reveal multiple photoresponse mechanisms based on the photovoltaic, photoconductive, and photothermal effects.

Ahn et al. [50] studied the individual Si NW field-effect transistors using SPCM for the first time. They reported a photoresponse of Si NW with and without voltage bias. When Si NWs are biased, the photoresponse depends highly on the incident light polarization angle with respect to the sample, and the photocurrent generation is centered in the middle of the Si NW between the contacts (***Figure 6*****a-b**). The photocurrent is ~110 nA at 100 kW/cm$^2$ laser power for the laser polarization aligned with the Si NW axis. They reported a very strong polarization dependence of the generated photocurrent and claimed that the anisotropic light absorption mechanisms are responsible for the polarization dependence of the photocurrent. A bipolar response is observed under zero bias at the opposing contact points of the nanowire with the nickel electrodes. Longer nanowires are used to illustrate the open-circuit photocurrent generation. The amplitude of the zero-bias photocurrent is ~10 nA at ~100 kW/cm$^2$ incident laser power (***Figure 6*****c**). They attribute the zero-bias photocurrent generation near the contacts to the separation of electron-hole pairs due to the bent electronic bands near the metallic contacts. The gate-dependent measurements are also performed in their study, and a dramatic change in the photocurrent is observed. Their simple band model adequately explains the changes in their gate-dependent photocurrent measurements.

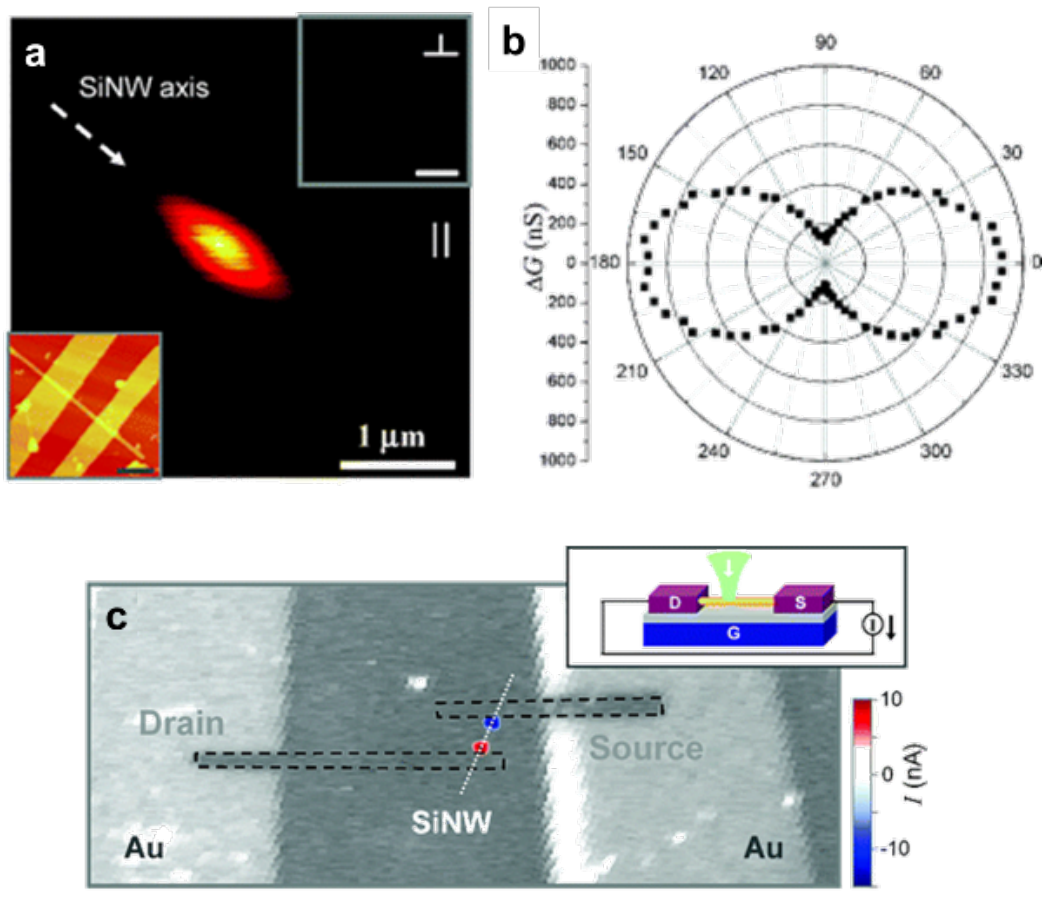

**Figure 6 a.** SPCM map of a Si NW under 100 mV bias with parallel polarization. The upper inset shows the SPCM map at perpendicular polarization, and the lower inset shows the AFM height trace map of the sample. **b.** Polarization-dependent photoconductance change is given. **c.** SPCM map overlayed on a reflection map is given. Response is localized near the contacts. The inset shows the SPCM configuration. Adapted with permission from Ahn et al. [50]. Copyright 2005 American Chemical Society.

Although they discuss the unlikelihood of the photothermal origin of the signal, we think that further investigation is required, as it is hard to draw a conclusion based on the experimental results reported. First, the laser power they used was quite intense. As a comparison, SPCM experiments on CNTs, with a much smaller optical absorption cross-section, showed a temperature increase large enough to induce photothermal effects. A similar laser-induced heating for Si NWs is plausible. The high-power-density spot used in Ahn et al. SPCM measurements might generate an emf due to the Seebeck effect. Bulk Si has a relatively large Seebeck coefficient of around -900 µV/K near room temperature [51]. Much larger Seebeck coefficients in the range of -1500 µV/K have been reported for arrays of lightly n-doped Si NWs by Krali et al. [52]. Yet, even for more conservative Seebeck coefficients, around -170 µV/K of CMOS-compatibly produced Si NWs [53], we estimated that a ~1 K temperature rise is sufficient to create the reported zero-bias photocurrent. In this estimation, we deduced the resistance from the conductance measurements reported in the paper. Of course, the photovoltaic response should also be taken into account, as both effects might be present simultaneously. However, it would be instructive to perform further SPCM experiments, accompanied by other methods, before concluding about the nature of the photoresponse.

In a similar study, Kelzenberg et al. [54] studied the photocurrent generation in a single Si-NW solar cell configuration using SPCM. They formed electrical contacts with Ag-capped Al to achieve Ohmic junctions to the Si NWs. Then, rectifying junctions are formed by applying extreme current between the two contacts at one end of the nanowire in a 4-wire configuration. This creates a rectifying junction on one end of the nanowire while the other remains ohmic (***Figure*** *7***a-e**). The fact that the photocurrent is only observed at the rectifying junction illustrates the presence of the photovoltaic response. [49]They extracted the minority carrier diffusion length as ~2 µm. This is almost two orders of magnitude larger than the measurements performed using electron beam-induced current in similar Si NWs [55]. Also, the significant asymmetry of the contact topography must be taken into consideration, as it may alter the absorbed laser power by the Si NW.

Triplett et al. performed SPCM over suspended Si NWs [56]. They extracted the open-circuit photocurrent profiles and claimed that the minority carrier diffusion length is suspended Si NW is ~2.7 µm. As a comparison, they fabricated supported devices and performed EBIC and SPCM to show a 45 nm minority carrier diffusion length (***Figure** 7***f-g**). However, EBIC studies are missing for the suspended devices in the paper. In the supporting information of the paper, they also modeled the temperature distribution along the nanowire; however, they ignored the Newtonian cooling and ended up with a conclusion stating that a linear temperature is expected along the NW. However, in very similar structures, other groups and we have shown that the temperature (and photocurrent) profile can deviate from linear due to Newtonian cooling [57,58].

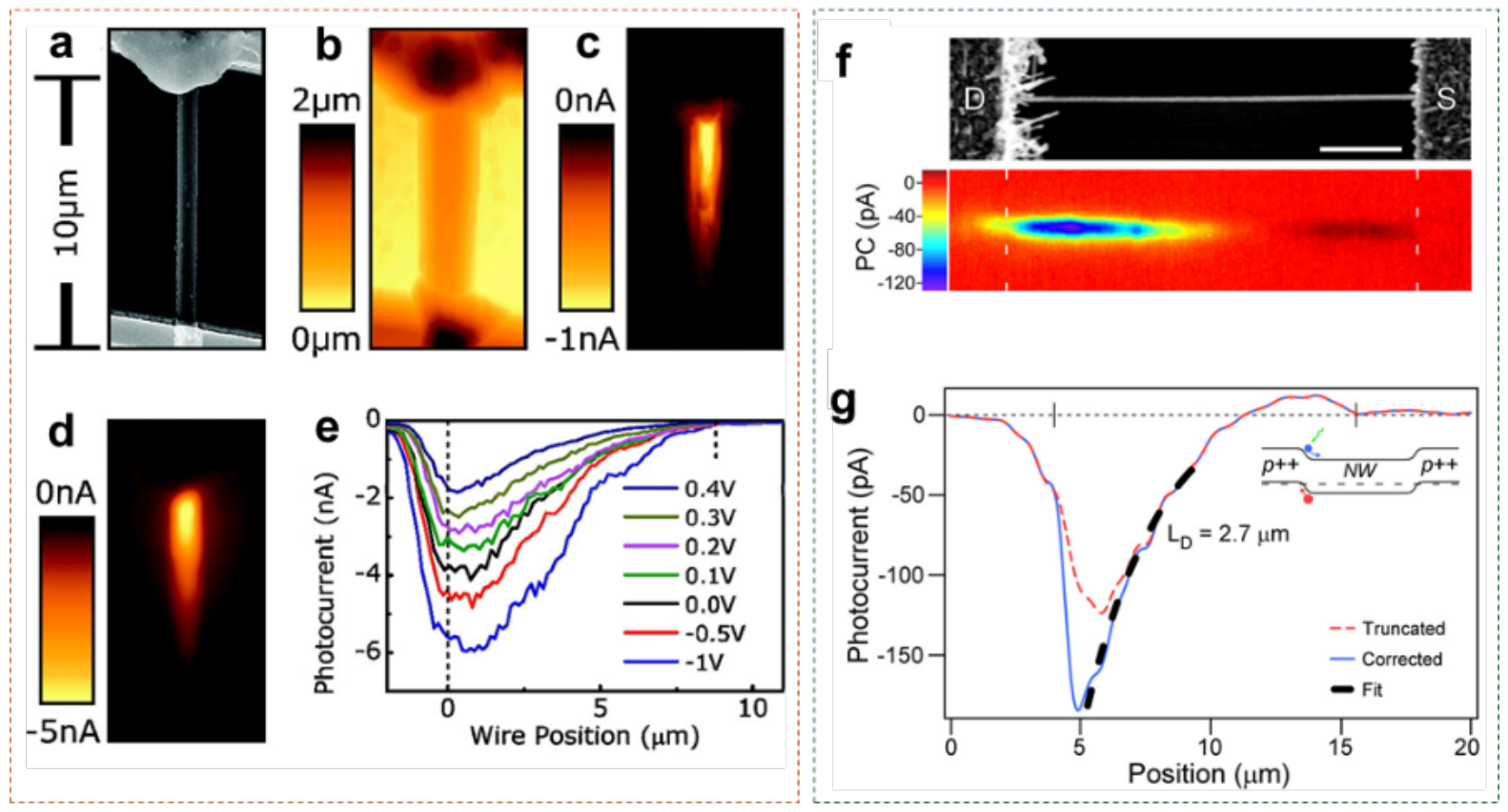

**Figure 7 a.** SEM image of Si NW device with silver contacts. **b-c.** NSOM height trace map and photocurrent maps, respectively. **d.** SPCM photocurrent map and **e.** line traces through the SPCM maps at different biases. Adapted with permission from Kelzenberg et al. [54]. Copyright 2005 American Chemical Society. **f.** SEM image and SPCM map of suspended Si NW devices. **g.** Line trace of photocurrent along the Si NW. Adapted with permission from Triplett et al. [56]. Copyright 2015 American Chemical Society.

Another interesting study on Si NWs has been performed by Mohite et al. [59]. They studied the photoresponse of a Si p-n junction NW with a diameter of <100 nm. They performed SPCM measurements using a 532 nm laser at 100 μW (20 kW/cm$^2$) and reported the photoresponse profiles along the nanowire under various biases. The photoresponse peaks around the p-n junction, and the photoresponse peak shifts slightly with the applied bias. On both sides of the p-n junction, the minority carrier lengths have been extracted as ~1 μm. However, again, it would be difficult to differentiate any possible photothermal contribution solely by SPCM with a laser above the band gap. The long minority carrier diffusion length conclusion can be misleading because of contributions from the photothermal effects to the total signal registered by the SPCM.

Allen et al. [60] studied vapor-liquid-solid (VLS)-grown n-Si nanowires with a 10 μm long region of the surface etched. As a result, they created an FET structure with improved mobility and large on/off ratios. They showed that the etched section of the NW exhibits a photoresponse with gate-tunable characteristics. Their gate-dependent studies show that within the NW region, at the subthreshold regime, large resistance suppresses the photoconductance from the NW. In other regimes, the resistance is low enough to measure photoconductance over the NW. Moreover, in a follow-up paper, Allen et al. [61] studied p-doped Si NWs with an etched device channel as in the previous study and did a quantitative analysis of the results. They demonstrated a resistivity gradient along the nanowire. These studies demonstrate the use of SPCM on resistive NW devices and how changes to device configurations can change the measured signal.

Before concluding the part on Si NWs, we would like to direct the reader's attention to the photothermal effects one more time. In a study by Roder et al. [62] Si NWs in an optical trap are shown to heat more than a few Kelvin in an aqueous environment at an irradiance of 1 MW/cm$^2$ at the wavelength of 975 nm. Although the irradiance is large compared to the SPCM measurements we discussed, so is the temperature increase of the Si NW. Such a temperature increase may result in significant photothermal effects along with photoresponse from photovoltaic and photoconductive origins. Also, it should be noted that the light absorption of the Si NW at the trapping laser wavelength is very small. At the wavelengths where the absorption is large, Si NWs should heat up more. On that note, we encourage people in the field to look further into any possible thermal effects, as this might be imperative in solar energy harvesting-related applications of Si NWs.

SPCM has been employed in studying many different semiconducting nanowires beyond CNTs and Si NWs. We will not discuss the details of the studies here; however, we will briefly mention it for the sake of not repeating the major contribution of SPCM in the study of semiconducting NWs. One interesting study by Yang et al. on PbS [52] NWs report important observations on the interpretation of SPCM results in lightly doped one-dimensional semiconductors. Since the dark majority carrier concentration is low, optically injected carriers can overwhelm the dark carriers and can significantly alter the band bending of the semiconducting NW near the contact electrodes. Moreover, such materials can be significantly influenced by the screening provided by metallic contacts, as they can attenuate the gate field and lead to a non-uniform carrier concentration along the nanowire near the contacts.

SPCM studies on Ge NWs reveal a diameter-dependent internal gain mechanism. Kim et al. [63] studied multiple two-terminal Ge NW devices fabricated on a tapered crystal. It is noteworthy that small-diameter sections exhibit larger photocurrent generation as compared to the large-diameter devices. They find that the relatively high photoconductance of narrower devices is attributed to hole accumulation near the Ge/GeOx interfaces due to the presence of surface-trapped electrons. This results in a diameter-dependent density of charge carriers as neatly demonstrated by a series of experiments and analytical modelling. Shin et al. used SPCM to measure the photocarrier diffusion length in intrinsic Ge NWs [64]. They measured the photocurrent profile beyond the source-drain contacts and fitted exponential functions to find the carrier diffusion lengths. Despite no consideration being given to other possible effects, this is a systematic way to extract diffusion lengths in semiconducting NWs.

### 2.3. Nanowires of Vanadium Dioxide as a prototypical correlated material

Nanowires of materials with strong electronic correlations also exhibit intriguing photocurrent generation. The case we would like to focus on is nanowires of $VO_2$. $VO_2$ is an intriguing material that exhibits various phase transitions from insulating to insulating and metallic phases depending on the pressure, temperature, and doping [65]. Upon the metal-insulator transition, a 0.6 eV energy gap opens due to the electronic correlation. Here, this is not a single-electron band gap as in conventional semiconductors since the quasiparticle picture fails when the electronic mean free path is comparable to the interatomic spacing. Moreover, the metallic phase of $VO_2$ above 65 °C is at the Ioffe-Regel limit of conductivity.

In a previous study, we used SPCM to elucidate the photoresponse in $VO_2$, especially when the metallic and the insulating phases are in coexistence [58]. There is a neat trick to achieve this condition by suspending a clamped $VO_2$ nanowire across the metal electrodes [66]. Soon before our publication was out, Miller et al. [67] studied two-terminal supported $VO_2$ nanowires using SPCM. Their analysis has been based on the single-electron picture, and this analysis showed unusually long carrier lifetimes in a correlated semiconductor. Despite the low-intensity laser power (~500 W/cm$^2$) they employed in their experiments, the temperature increase on the crystal could still be on the order of 50 mK. The Seebeck coefficient difference between the insulating and the metallic phase is on the order of ~-0.5 mV/K [68]. Thus, when the nanobeam is in coexistence around 60 °C the emf generated due to the Seebeck effect is on the order of 0.05 mV. For a typical nanowire, the resistance is ~100 kΩ as in Figure 1(c) of Ref. [67]. A simple calculation yields a photoresponse on the order of 250 pA, which is larger than the value reported in the paper. Indeed, the built-in electric field-induced non-equilibrium charge carrier treatment of the observed response yields unrealistically long carrier lifetimes, which is highly unlikely for a correlated electron system.

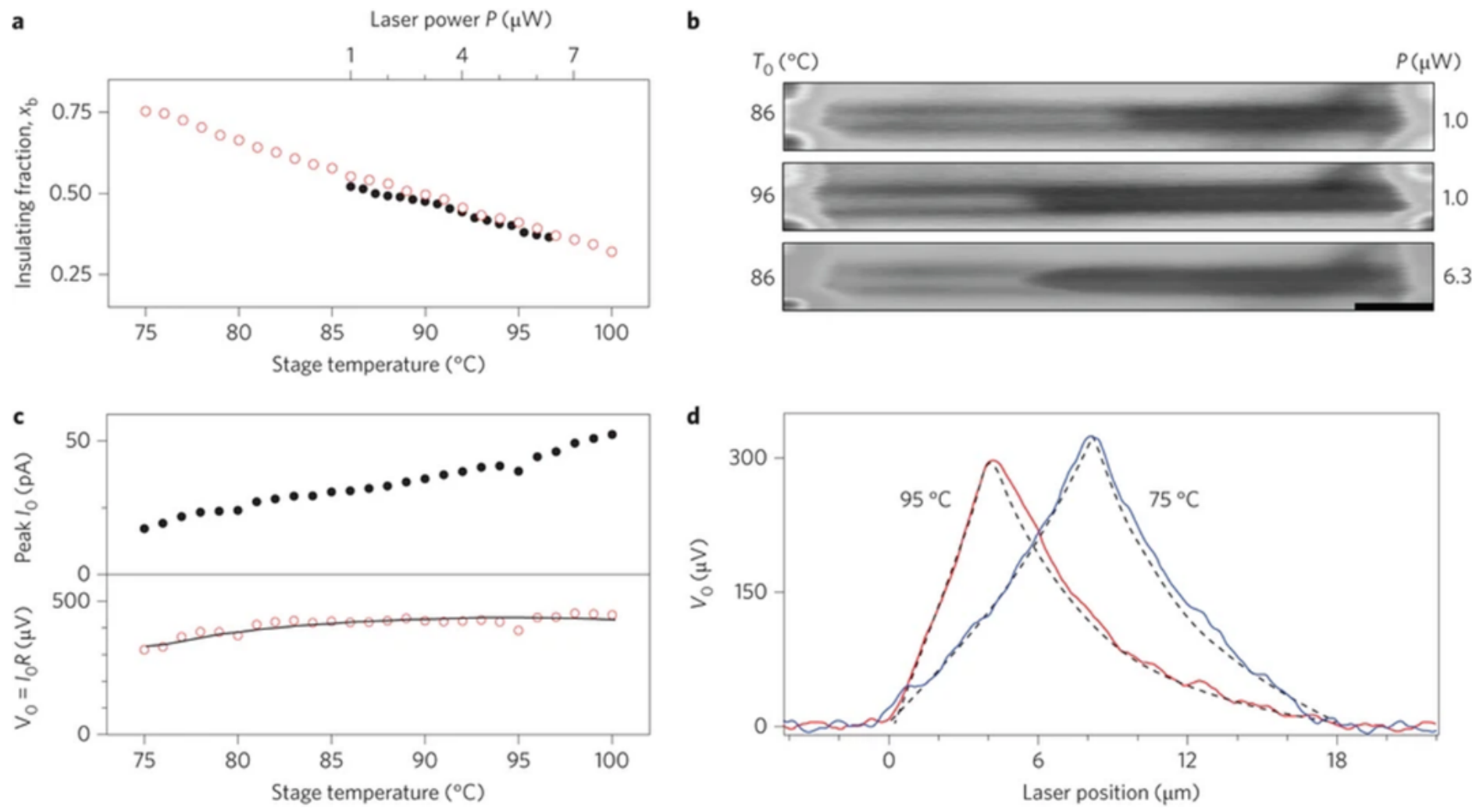

**Figure 8 a.** In the coexistence of suspended $VO_2$ nanobeams, the insulating fraction against stage temperature and the laser power shows a similar trend. **b.** SPCM reflection maps collected at different stage temperatures and laser powers show the effect of laser heating on the metal-insulator phase boundary. **c.** As the resistance of the sample decreases with the increasing metallic fraction, the peak photocurrent increases. This can be modelled into photovoltage as given in the lower panel. **d.** The photovoltage line trace along the crystal (solid lines) and the thermal model (dashed lines) show the photothermal origin of the response. Newtonian cooling is incorporated into the model. Adapted with permission from Kasirga et al. [58]. Copyright 2012 Springer Nature Limited.

In our experiments [58], as we suspended the crystals, the heating effect was unambiguously illustrated as the phase boundary was displaced under the very small laser powers (***Figure 8***). Likewise, when $VO_2$ nanobeam devices are biased, the photoresponse can solely be explained by the bolometric response. In a more recent study, Wang et al. [69] claimed that by tuning the laser chopping frequency (used for the reference of the LIA) in a wide-illumination photocurrent microscopy setup, they could distinguish the photothermal effects from the photo-injection. They claimed the photothermal process could be suppressed, especially above 2 kHz chopping frequency. This is a rather confusing claim, as the ultrafast pump-probe spectroscopy measurements with below-threshold fluence show a response in the ps timescale for the heat diffusion duration [70]. In our opinion, a more likely explanation for their observation is that, on the kHz time scale, due to the decrease of the actual gain of the pre-amplifier vs. the nominal gain, they observed a decrease in the photoresponse. Eliminating the pre-amplifier should bring back the suppressed signal. This is particularly important when very high sensitivities are employed.

### 2.4. Metallic Nanowires

As we introduced earlier, it is possible to get a photoresponse from metallic nanowires in an SPCM measurement. The mechanism is based on the laser beam-induced heating. This becomes particularly significant in suspended nanowires, such as in the case of metallic CNTs discussed in Subsection 2.1 or metallic $VO_2$ crystals as discussed in Subsection 2.3. Recently, we studied the photoresponse of single and networked Silver NWs [20]. Our detailed measurements and analysis show that the bolometric effect is responsible for the observed photoresponse. Moreover, when two nanowires form a junction, the incident beam can cause a plasmonic field enhancement at the junction due to the separation caused by the polyvinyl alcohol (PVA) surfactant. As a result, temperatures much higher than what can be induced on a single nanowire with the same laser power are realized. This temperature increase causes local resistivity to increase dramatically, leading to the observed bolometric photoresponse.

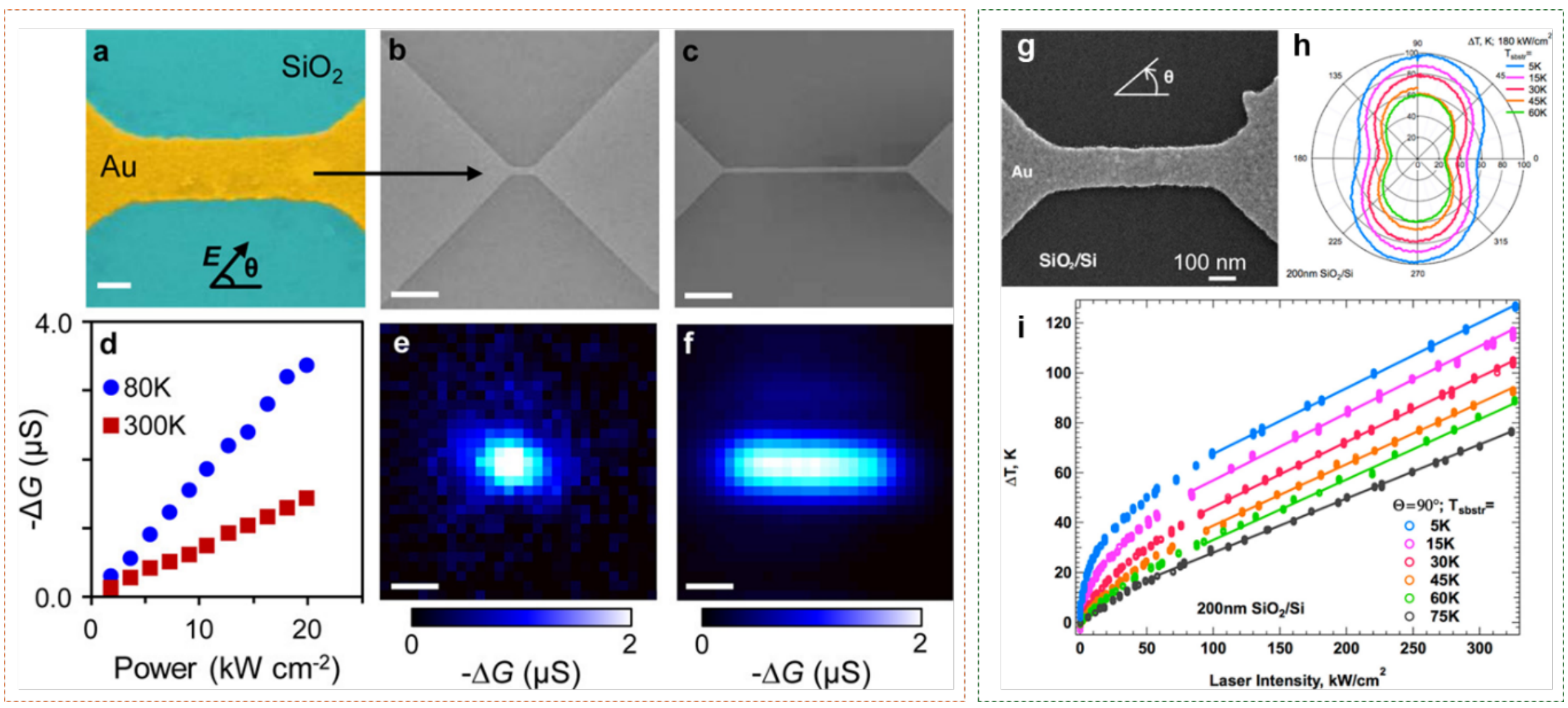

**Figure 9 a.** False colored SEM image of an Au NW device on $SiO_2$ with polarization angle indicated. **b.** and **c.** show SEM images of Au NW devices with different lengths. **d.** Photoconductance change vs. laser power at 80 and 300 K. **e** and **f.** show SPCM photoconductance change map from the devices. Adapted with permission from Herzog et al. [21]. Copyright 2014 American Chemical Society. **g.** SEM image of an Au NW device. **h.** Polarization dependence of the measured temperature change in the NW at different global temperatures. **i.** Laser intensity vs. sample temperature change at different global temperatures with the polarization of the incident beam aligned with the NW axis. Adapted with permission from Zolotavin et al. [71]. Copyright 2016 American Chemical Society.

In a similar study, Herzog et al. [21] performed SPCM on lithographically defined Au NWs. Their work shows that resonant and non-resonant plasmonic effects lead to a large bolometric response. As a result, the change in the resistivity leads to a large negative photoconductance. Light intensity of 20 kW $cm^{-2}$ leads to a temperature increase of 4.5 K for 100-180 nm wide Au NWs. Photothermoelectric effects in Au NWs with nanogaps are also shown to become dominant, especially when the local Seebeck coefficient is altered due to the nanowire width [72]. In a recent study, Zhao et al. [73] investigated the photonic contributions to the Seebeck photovoltage generation of metallic Au nanowires using SPCM. A stripe of Au NW of 400 nm width is coupled to plasmonic structures, gold discs of diameters ranging from 400 to 1000 nm. At a large laser power intensity (~10 kW $cm^{-2}$ at 532 nm), regions with plasmonic structures exhibit a thermoelectric response at the edges of the regions with plasmonic structures. Although the exact mechanism is unclear in their studies, various candidates, such as alteration of the Seebeck coefficient of the nanowire due to plasmonic structures, width variations due to the fabrication procedure, are likely candidates for the observed response. Regardless of the physical origin, this work sets a very good example of how substrate or environment modification can lead to photovoltage response in low-dimensional materials. It is also noteworthy that slight width alterations in gold nanowires can result in variations in the Seebeck coefficient that can be used for a single-metal photodetector [74]. Remote heating from surface plasmon polaritons (SPPs) is also important in metallic nanowires, as quantified by SPCM studies. Evans et al. [22] demonstrated that gratings at the contacts of metallic NWs can couple SPPs for remote heating of the NW. Our SPCM studies also showed similar results in Ag NWs and their networks [20].

In another study [71] by the Natelson group, SPCM experiments on Au NWs have been performed at low temperatures to elucidate the role of thermal boundary resistance. They observed a nonlinear dependence of bolometric response on the incident laser intensity. This is attributed to the temperature-dependent alterations to the thermal boundary conductance between the NW and the substrate. This study demonstrates the role of laser heating and the substrate effects on SPCM measurements. Despite the fact that the signal is enhanced by plasmonic effects, this is not the only contribution to the observed signal. **Figure 9** summarizes the results from these two important studies. As we will discuss in detail in Section 3, these results are also relevant to the SPCM results from two-dimensional materials.

## 3. SPCM on two-dimensional materials

Atomically thin materials provided a route to study many solid-state problems with relative simplicity. Unlike the complicated and costly epitaxial deposition of the semiconducting heterostructures, both mechanical exfoliation and chemical vapor deposition methods can be conveniently employed by many laboratories to obtain 2D materials. Moreover, unlike most two-dimensional electrons at the interfaces of doped semiconductors, the electrons in 2D materials are optically accessible. At the time this review is written, hundreds of atomically thin materials have already been studied experimentally in the literature, and hundreds more are waiting to be synthesized.

The first SPCM studies on 2D materials have been performed on graphene [75]. Various studies illustrated the photoresponse of graphene pn junctions or interfaces and elucidated the underlying mechanisms [10,18,19,26,75–80]. We will discuss the role of the SPCM in exploring the underlying photoresponse mechanism. $MoS_2$ monolayer also attracted considerable interest thanks to its direct band gap [81]. SPCM has been employed in several studies to elucidate the underlying mechanisms of the photoresponse in $MoS_2$ mono and few layers [31,82,83]. SPCM has been used in many other 2D materials, including semimetallic and metallic ones. In a recent study, we also demonstrated that, besides the fundamental investigation of the photoresponse generation, an SPCM-based method could be employed to measure the thermal conductivity of atomically thin materials [27,84]. The following subsections give an overview of these topics.

### 3.1. Photoresponse mechanisms of graphene determined via SPCM

Graphene is a single layer of graphite made up of a hexagonal arrangement of carbon atoms in sp2 hybridization [85]. The peculiar band structure of graphene leads to a pseudo-relativistic description of the low-energy excitations [86]. The absence of an energy gap at the linear dispersion relation near the K point results in a fascinating photocurrent response. Moreover, peculiar electron-phonon scattering pathways result in a bottleneck in the transfer of thermal energy to the lattice [19,87]. As a result, the intrinsic photoresponse in graphene results from the hot electrons, as elucidated by various SPCM-based studies [10,18,76,80].

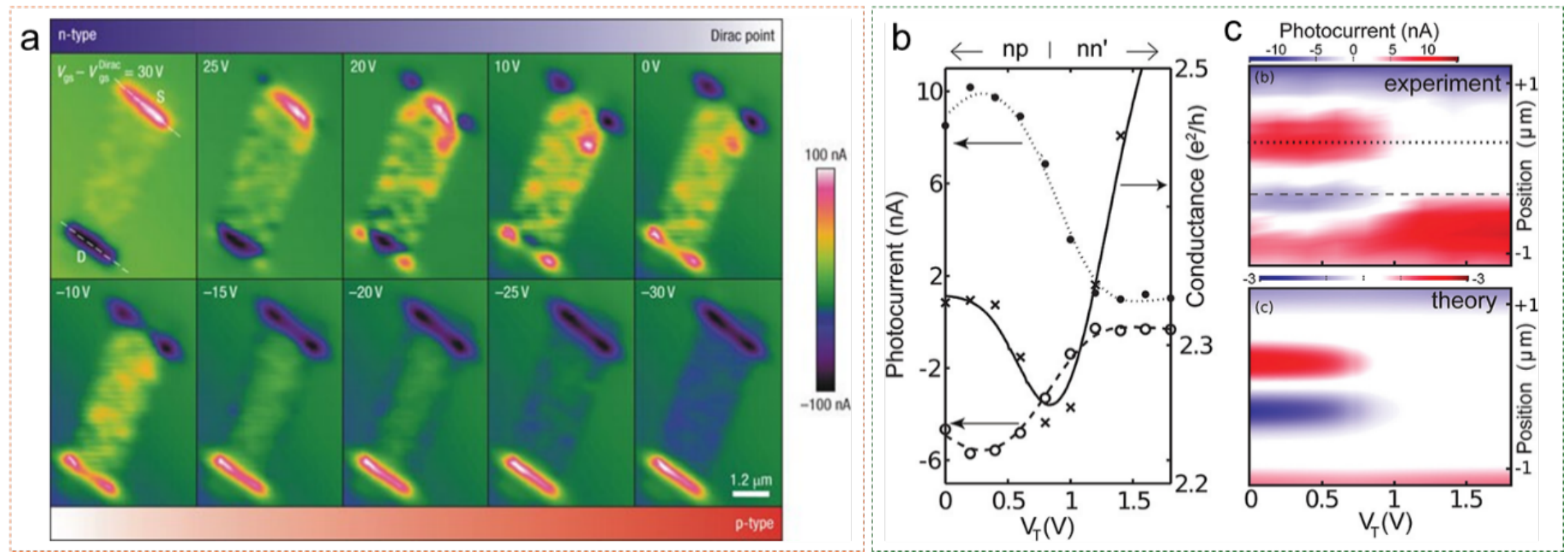

**Figure 10 a.** Photocurrent maps of a monolayer graphene device under various gate voltages. Adapted with permission from Lee et a [75]. Copyright 2008 Springer Nature Limited. **b.** Photocurrent as a function of top gate voltage with the laser positioned at either side of the top contact. **c.** The upper panel shows the line trace across the device versus different top gates around the contacts. The lower panel shows the theoretical model. Copied with permission from Lemme et al. [80]. Copyright 2011 American Chemical Society.

Many SPCM studies have been performed on graphene and its multilayers to understand their photoresponse mechanisms. One of the first reports on employing SPCM on graphene was by Lee et al. [75]. In 2008, they observed photoresponse both near the contacts and over the graphene flake. The response near the contacts is attributed to the built-in electric fields that induce a separation of non-equilibrium charge carriers, and the latter is attributed to the spatial potential fluctuations associated with the formation of electron-hole puddles. They also demonstrated that the photocurrent profile can be altered by applying a gate voltage through the back gate (**Figure 10a**). Lemme et al. [80] also used SPCM on monolayer graphene and showed that a strong photocurrent is generated near the contact electrodes (even if they are floating). The observed response is attributed to the electron-hole separation due to the built-in electric fields near the contacts. However, when the flat band is achieved via gate voltage, the photoresponse diminishes. They also demonstrated that thermoelectric currents near the metal contacts are present in the flat band condition (**Figure 10b-c**).

In a 2009 paper, Mueller et al. [88] used a scanning near-field photocurrent microscopy (SNPCM) setup to study monolayer graphene and monolayer/multilayer graphene junctions. By using the high spatial resolution of the SNPCM, they demonstrated that the electronic structure modification provided by the metal contacts extends several hundred nanometers into the graphene. In 2010, Peters et al. [76] published their findings on the photoresponse of a graphene p-n junction revealed by SPCM. The graphene p-n junction is formed via chemical doping of one-half of the flake using polyethyleneimine solution. Moreover, they applied a gate voltage to globally control the carrier density of the sample. Their measurements showed that the signal is dominated by the built-in electric potential across the p-n junction, yet the photothermoelectric contribution to the observed signal cannot be ruled out.

Although chemically doped p-n junctions studied via SPCM provided insight into the photoresponse mechanisms in 2011 and 2012, various SPCM studies on graphene p-n junctions formed via electrical gating have been published. First, Lemme et al. [80] fabricated a device with a top and bottom gate. The bottom gate was used as a global gate, and the top gate was a finger gate isolated from the graphene layer via an $Al_2O_3$ dielectric. As a result, by controlling the back gate, they could tune the device from n-p-n to p-n-p to n-$n^-$-n to p-$p^+$-p doping. They demonstrated that, contrary to the earlier study by Peters et al. [76] the photothermoelectric (PTE) effect dominates the photoresponse. It is important to note that they used "modest" laser powers in their experiments (i.e., 0.5 μm wide spot at 40 μW, 600 nm wavelength). They show that for a Seebeck coefficient [89] of ~50 μV/K, a temperature increase of ~1 K is sufficient to induce the observed photoresponse. Later, these findings are also supported by a theoretical analysis by Song et al. [19].

Final experimental evidence that the photoresponse in a graphene p-n junction is dominated by the hot-carrier-assisted photoresponse is shown by Gabor et al. [18] and Sun et al. [10]. Both studies used SPCM, with the latter incorporating time-resolved photocurrent measurements to elucidate the time scale of the observed response. Top-gated monolayer graphene devices with slightly different geometries are used in both studies. In graphene, photoexcited carriers initially relax via electron-electron scattering and optical phonon emission, followed by a novel transport regime where further relaxation of electronic energy to the lattice is quenched [19,87,90]. During this period, the photogenerated charge carriers remain hot, and transport is performed via these hot carriers. Both studies experimentally show that when the top gate is swept, there are multiple zero-photocurrent crossings, while the photovoltaic contribution could only give a single zero-photocurrent without the PTE contribution.

Before concluding the review of SPCM studies on graphene, we would like to mention the study by Cao et al. [91] where they showed that under a magnetic field, the photo-thermal gradient created by the diffraction-limited laser spot leads to a Nernst current at the opposing edges of the two-terminal graphene device. The results show photo-Nernst effect can be detected with magnetic fields as small as 50 mT in graphene. This suggests that SPCM can be used in other topological matter to investigate quasiparticle properties of materials with short scattering times.

The role of SPCM has been crucial in elucidating the photoresponse mechanisms in graphene. Unconventional nonequilibrium carrier relaxation mechanisms in the quantum-confined electronic system lead to a peculiar photoresponse that could potentially be useful in various optoelectronic applications. Next, we focus on the use of SPCM in semiconducting two-dimensional materials.

## 3.2. Photoresponse of semiconducting 2D materials determined by SPCM

There are many semiconducting 2d materials. Here, we will focus on the use of SPCM in the photoresponse of the most widely studied materials in the TMDC family and black phosphorus. We will try to be as comprehensive as possible to provide a general overview of how SPCM has been utilized in these studies and the interpretation of the measurement results.

### 3.2.1. Use of SPCM on $MoS_2$ and other transition metal dichalcogenides

$MoS_2$ is one of the first examples to be exfoliated down to a single layer following the advent of graphene [81]. When exfoliated to a monolayer, $MoS_2$ becomes a direct gap semiconductor with an energy gap of 1.8 eV [81]. Such a transition to a direct gap makes $MoS_2$ an optically exciting and relevant material. $MoS_2$ is an archetypal material, and its close relatives, $MoSe_2$, $WS_2$, and $WSe_2$, also exhibit very similar photocurrent mechanisms. Although we will discuss some differences in the photocurrent generation among semiconducting TMDCs, our focus will be on the mono- and few-layer $MoS_2$.

Photocurrent generation is prominent in both monolayer [92,93] and multilayer [94] $MoS_2$. The first reports mainly focus on broad illumination of the phototransistor structures composed of a Si back gate with $SiO_2$ as the gate dielectric. Yin et al. [92] studied mechanically exfoliated monolayer $MoS_2$ device structure and demonstrated photoswitching with 50 ms rise and fall time at room temperature in air. The transistor structure is also used to modulate the photoresponsivity of the device. The photoswitching and its gate dependence are explained by the transport of photoexcited carriers. In a follow-up study by Zhang et al. [93] monolayer CVD crystals are studied in the Si back-gated phototransistor structure, both in air and under high vacuum. Their results show a persistent photoconductance, a state of high conductance even after the excitation is removed, both in air and vacuum. Both the reported responsivities and rise/fall times are drastically different from the Reference [92]. Besides the synthesis method, the crystals Zhang et al. used were polycrystalline. Later studies on the effect of chalcogen vacancies and ambient moisture shed light on the results of Zhang et al.

The first SPCM studies in the literature on $MoS_2$ monolayers and few layers were reported by Buscema et al. and Wu et al. in 2013, respectively [31,82]. Buscema et al. [31] demonstrated that the photoresponse at the junction between the monolayer $MoS_2$ and the Au/Ti electrode is dominated by a large and tunable thermoelectric effect. They employed two different laser wavelengths, one at 532 nm (2.33 eV) and the other at 750 nm (1.65 eV), with photon energies larger and smaller than the bandgap, respectively (**Figure 11a-e**). In both cases, the photoresponse is extended to the contacts and is very symmetric with respect to the metal-semiconductor junction. This shows that the photovoltaic effect due to the formation of the Schottky barrier is not mainly responsible for the generated photoresponse. Moreover, they used the Si substrate as the back-gate to modulate the observed photoresponse by tuning the Seebeck coefficient of $MoS_2$ via gating. The laser power levels in the study are 1-60 μW. Considering the laser spot size of ~400 nm, the laser irradiance is 200-12000 W $cm^{-2}$. Even for such low laser powers, the authors estimated the temperature rise at the Au/$MoS_2$ junction to be 0.04 to 0.4 K, sufficient to induce a photothermoelectric response.

Wu et al. [82] performed a similar study on a few-layer $MoS_2$ transistor structure. To gain a deeper understanding of the mechanisms leading to the photoresponse, they performed time-resolved SPCM measurements while controlling the source-drain and gate bias. They concluded that for 4-layer $MoS_2$, the photoresponse is dominated by diffusion and drift of photoexcited carriers in the regions of high electric field, both in the device channel and in the vicinity of the contact/$MoS_2$ junctions. They performed SPCM measurements with 20 μW (4 kW/cm$^2$) excitation power. The photocurrent spectra at this constant power show an increased photocurrent generation of around 650 nm, for single, 3, and 4 layers of $MoS_2$ (**Figure 11f-g**). It is unclear which part of the sample the photocurrent is measured from. They attribute the increase in the photoresponse to the increased photoexcited charge carriers due to the interband absorption. Based on this measurement, they rule out the possibility of carrier generation in or heating of the metal contacts. However, two aspects must be further investigated to fully reach the conclusions in the study, in our opinion. One is the wavelength-dependent thin-film interference from the 300 nm $SiO_2$ / $MoS_2$ interface, and the other is wavelength-dependent absorption changes of the $MoS_2$. Although the authors touch upon both mechanisms in the manuscript, the lack of experimental evidence raises questions regarding the conclusions they reach. Moreover, our recent studies [95] show that, by simply varying the depth of the trench on which the $MoS_2$ is suspended (analogous to varying the wavelength for thin-film interference), the absorption can be significantly varied.

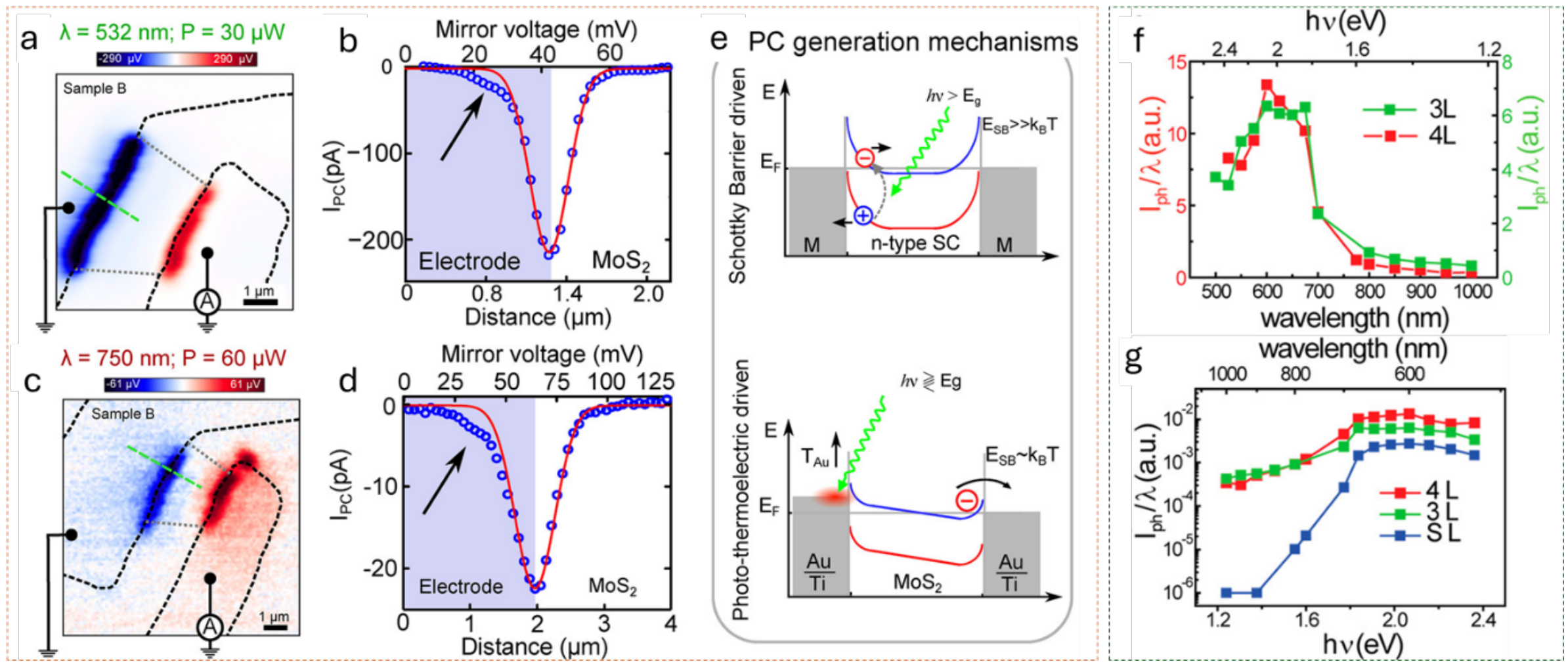

**Figure 11 a.** SPCM map of a $MoS_2$ monolayer device collected with a 532 nm, 30 μW laser. The electrode edges are marked with dashed lines. The green dashed line indicates the line trace shown in **b.** A similar set of data collected on the same device with 750 nm is shown in **c** and **d. e.** Photocurrent generation mechanisms are depicted. Adapted with permission of Buscema et al. [31] Copyright 2013 American Chemical Society. **f.** Normalized photocurrent vs. excitation wavelength for 3 and 4 layers of $MoS_2$ is shown. The photocurrent is normalized with the wavelength to account for the changes in the photon flux at different wavelengths. **g.** Comparison of spectra for monolayer, 3, and 4 layers on a semilog scale. Adapted with permission from Wu et al. [82] .Copyright 2013 American Chemical Society.

Wu et al. also performed bias-dependent SPCM measurements on their $MoS_2$ devices. They demonstrated that depending on the sign of the bias, the photocurrent is enhanced in the vicinity of one of the contacts while suppressed near the other. They explained this observation by the band bending near the contacts. Despite the deductions presented in the paper, we think there is not enough evidence to fully exclude the photothermal effects. The photothermoelectric effect, in conjunction with the photovoltaic effect, could result in a similar response. The extent to which it is dominant under what conditions should be complemented by another method that doesn't involve light, such as scanning thermal microscopy or SPCM with excitation energy smaller than the band gap.

In a more recent study, Jiang et al. [83] used SPCM to demonstrate the flexo-photovoltaic effect in multi-layered $MoS_2$. The authors claim that a strain gradient is induced by placing a 16-layer-thick $MoS_2$ at the edge of a 118 nm thick $VO_2$ crystal. Then the authors claim that by heating the heterostructure, the strain gradient can be tuned, especially above 65 °C, and the metal-insulator phase transition in $VO_2$ leads to a change. Despite the intriguing interpretation of the results, as we have seen in other studies, the partial suspension of $MoS_2$ in the vicinity of the tall $VO_2$ crystal could lead to significant alterations in the absorbed laser power. Moreover, such suspension can lead to a significantly increased local temperature, which can result in the currents observed in the study. Similarly, the polarization dependence of the signal should be carefully investigated as the presence of $VO_2$, a conductive material, could alter the electrical field position and intensity, especially at high temperatures, above the MIT temperature.

In another study, 3R-$MoS_2$ was studied by Dong et al. [96] using SPCM to demonstrate the giant bulk piezophotovoltaic effect. The effect emerges from the built-in polarization field of the non-centrosymmetric 3R-$MoS_2$ crystal. Moreover, the crystals are strained to enhance the built-in polarization field. SPCM shows that when the strained crystal is oriented along the armchair direction, the polarization field points perpendicular to the edge of the contacts, and the photoresponse differs from that of the zigzag-oriented crystal. One of the points not considered in the study is the changes in optical absorption due to the differences in the distance between the substrate and the crystal. This is very important as in certain cases, the destructive interference of the incident laser with the light reflected from the substrate leads to a dramatic decrease in the measured photoresponse.

Despite the fact that there are many studies on materials like $MoSe_2$ [97–101], $WS_2$ [102–105], and $WSe_2$ [106–112], for the sake of brevity, we will not discuss them here. Most studies have conclusions similar to those discussed on $MoS_2$. One study we would like to mention is by Lu et al. [113], where they showed that laser irradiation could passivate chalcogen vacancies with oxygen in CVD-grown monolayer $WSe_2$ single crystals. They performed SPCM to show how laser irradiation alters the spatial distribution of the photocurrent. They irradiate the sample with a 300 mW laser at 532 nm, focused to a 1 μm spot via raster scanning over the sample. Laser irradiation leads to a dramatic change in the photocurrent distribution over the crystal. This dramatic change is attributed to increased free carrier lifetime and elimination of the traps at the interfaces. However, such extreme laser power is sufficient to oxidize the $WSe_2$, and such broad photoresponse can be attributed to the bolometric response [114]. Further experimental evidence is lacking in the report to show that the $WSe_2$ crystal in the SPCM study remained intact.

We would like to discuss the study by Furchi et al. [30] despite the study doesn't utilize SPCM, we found the results relevant to elucidating SPCM results from other studies reviewed here. The authors show that photovoltaic and photoconductive effects in the transistor structure are responsible for the photoresponse for the nW-level laser power impinging on the samples. Moreover, as the laser power increases, the responsivity of the sample decreases by more than two orders of magnitude, attributed to the photoconductive gain. In summary, the photoresponse of $MoS_2$ is due to a complex interplay of multiple effects. For a complete understanding of the underlying mechanisms at play for the given optical power, various mechanisms must be considered and systematically tested using methods complementary to SPCM. Such understanding is recently provided by Xu et al. [115] where they employed a scanning near-field microscopy-based SPCM to differentiate between various photoresponse mechanisms in $MoS_2$. Their setup allowed defocusing the tip to switch between photovoltaic and photothermoelectric effects (PTE) and they demonstrated that PTE extends several hundred nanometers into the $MoS_2$ from the gold contact edge.

**Use of SPCM in the valley physics of TMDCs**

Before concluding this part, we would like to discuss the application of SPCM in elucidating the valley Hall effect in TMDCs. In solids with broken time-reversal symmetry, the charge carrier in an electronic band gains a velocity perpendicular to the applied external field, which leads to the anomalous Hall effect (AHE) [116]. In semiconducting monolayer TMDCs, despite the broken inversion symmetry, this effect can be observed by creating a population imbalance between the two valleys (**K** and $\mathbf{K}^-$). To differentiate it from AHE, this effect is called the valley Hall effect (VHE) [117–119]. One way to create the aforementioned population imbalance between the two valleys is via using optical selection rules. Namely, depending on the helicity of the light, either of the valleys can be populated. As a result, VHE can be observed.

To elucidate the microscopic origin of the VHE SPCM is employed in various studies. Ubrig et al. [120] showed that photogeneration of both excitons and trions contributes to the occurrence of the VHE in monolayer $WS_2$ and bilayer 3R-$MoS_2$ crystals. Their results show that VHE is not mediated by interband transitions generating independent electrons and holes. Similarly, in a later study by the Heinz group [121], SPCM (just for the line traces) has been used to elucidate the response of the valley polarization of excitons under an out-of-plane magnetic field. Longitudinal photocurrent is modulated by the presence of the magnetic field, and this effect is attributed to the unbalanced transport of valley-polarized trions induced by the opposite Zeeman shifts of **K** and **K**' valleys. Beyond the VHE, valley-dependent optical selection rules and spin-valley locking can be used to achieve spin-polarized photocurrent[124]. In a monolayer $WS_2$ device with a spin-valve structure, Xie et al. [122] by employing SPCM demonstrated photocurrent with a high degree of spin-polarization. Similar results using SPCM have also been reported by Chen et al. [123]. All the aforementioned studies demonstrated the VHE at cryogenic temperatures. Li et al. [124] used chiral nanocrescent plasmonic antennae to selectively generate valley-polarized carriers in $MoS_2$ through hot electron injection, and used SPCM to map valley current in their four-terminal devices.

### 3.2.2. Other semiconducting 2d materials

We also would like to review SPCM studies on black phosphorus (BP), as it is an interesting material to study via SPCM due to its large anisotropy and tunable bandgap [125–127]. The inherent structural anisotropy of BP emerges from the puckered structure of the lattice. The band gap depends on the number of layers and ranges from 2 eV in the monolayer to 0.3 eV in the bulk [128].

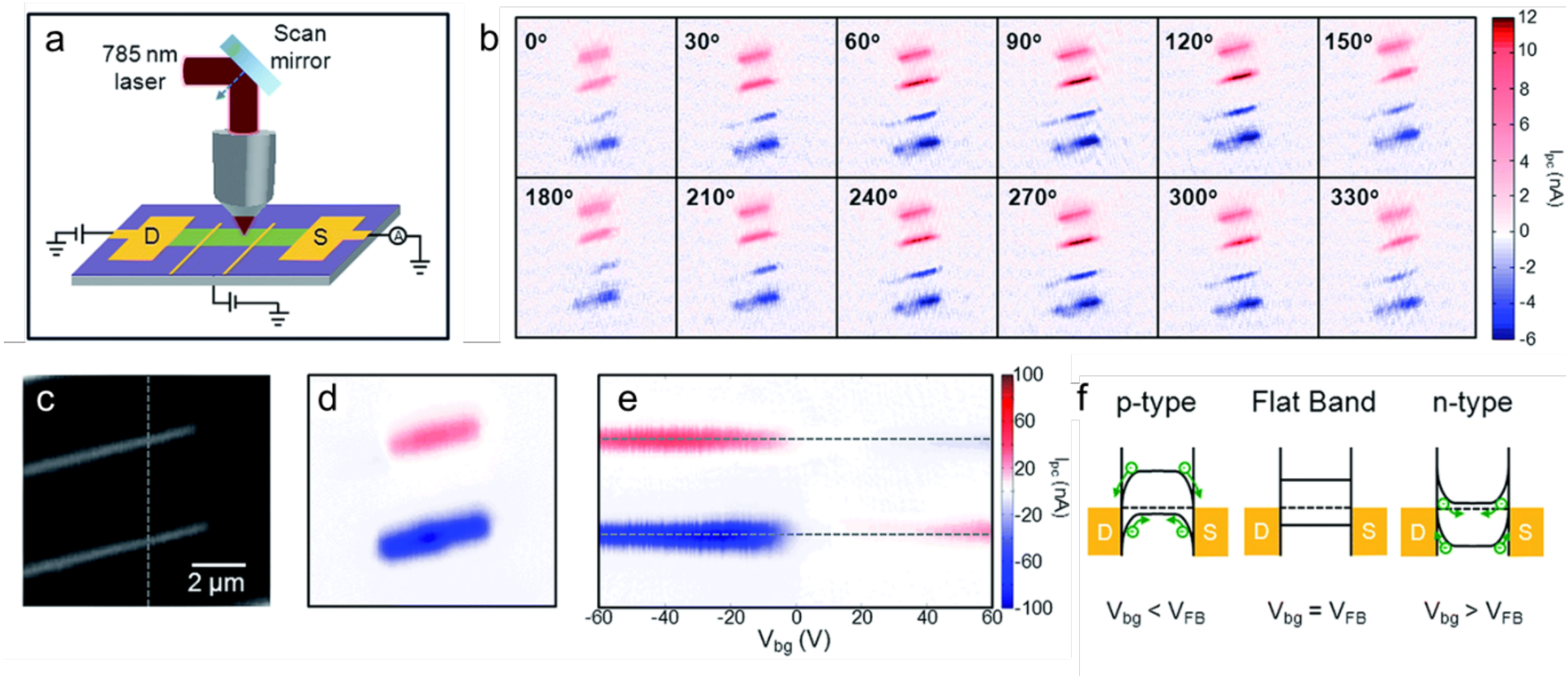

**Figure 12 a.** Schematic of the experimental setup and the black phosphorous device contact geometry. **b.** Polarization angle dependent photocurrent map under zero gate and sample bias, **c.** reflection map, and **d.** photocurrent map under zero gate voltage. **e.** Back gate voltage versus photocurrent line scan across the contacts. Each line represents a different gate voltage. **f.** Depiction of the band alignment depending on different gate voltages. Adapted with permission from Hong et al. Copyright 2014 Royal Society of Chemistry.

In one of the earliest SPCM studies on BP, Hong et al. [129] fabricated two-terminal devices with an extra pair of floating electrodes placed over 5-15 nm thick mechanically exfoliated BP crystals, as shown in **Figure 12a**. The floating electrodes are not electrically connected to a bias source or ground. A 290 nm $SiO_2$-coated Si substrate is used for gate-dependent measurements. SPCM measurements are performed under high vacuum with a polarized 785 nm (1.58 eV) laser. Under zero bias, samples exhibit bipolar photocurrent in the vicinity of the contacts (**Figure 12b**). The photoresponse has a slight polarization dependence, which is attributed to the anisotropy of the material. Moreover, gate-dependent SPCM measurements are performed near the inner floating electrodes under zero bias. The photocurrent exhibits modulation under gate voltage, and the polarity can be reversed at positive back gate voltages. The response is mainly attributed to the photovoltaic effect. The non-monotonicity of the response is further explained by the photothermoelectric contribution to the photoreponse due to the Seebeck effect. They also performed SPCM measurement when the samples were biased. However, due to the large contact resistance of the samples, the analysis of the data becomes difficult.

Hong et al. [130] performed SPCM measurements on four-terminal $MoS_2$-BP junction devices to elucidate the photoresponse mechanisms. 4.8 nm thick $MoS_2$ and 10 nm thick BP are used for the device fabrication. A 785 nm laser is used for the scanning measurements. No laser power figure is provided in the study for the SPCM measurements. SPCM maps under zero gate voltage show a 10 nA signal from the BP-$MoS_2$ junction. Pointwise photocurrent graph versus the gate voltage shows a modulable signal reaching up to 40 nA. The authors attribute the photoresponse to the photovoltaic effect. Moreover, polarization-dependent SPCM is performed using 532 and 1550 nm lasers at 1 and 35 μW power. Depending on the laser wavelength, the polarization dependence of the photocurrent at the BP-$MoS_2$ junction differs. While at 532 nm, the maximum signal is obtained around 150°, at 1550 nm, the maximum signal is obtained around 75° (and its 180° complement). Their Raman measurement shows that the polarization-dependence of the photocurrent is not dominated by the absorption anisotropy of BP. Rather, the difference is explained by the difference in photocurrent generation mechanisms depending on the excitation wavelength. The authors identify two different pathways depending on the laser wavelength. At shorter wavelengths, excitons can be created in $MoS_2$, in contrast to 1550 nm. In the latter case, anisotropy of BP dominates the signal, whereas under 532 nm excitation, photogenerated holes drift from $MoS_2$ to BP. However, as we discuss later in the case of $Bi_2Se_3$, polarization-dependent measurements have to be very carefully investigated, as the electric field continuity near conductive surfaces leads to changes in the photoresponse depending on the incident polarization. Considering the large and tunable Seebeck response of $MoS_2$ and the earlier work by the same group, further studies are required to fully elucidate the underlying response and polarization dependence.

Another 2014 study by Engel et al. [131] employed SPCM both at 532 nm and 1550 nm wavelengths on a 120 nm thick BP with multi-terminals. However, the focus of the study had been to demonstrate the use of BP as a photodetector at the aforementioned wavelengths. As a result, a little consideration is given to the underlying photoresponse mechanisms and the interpretation of the SPCM results.

A more comprehensive polarization-sensitive photoresponse study using SPCM is published by Yuan et al. [132] in 2015. A strong polarization dependence of the absorption of BP for a wide range of wavelengths is shown in the study. SPCM is performed both on a Hall-bar patterned and a ring-shaped metal electrode patterned BP crystals, ranging from 30 to 50 nm in thickness. In Hall-bar-shaped samples, a 100 mV bias was applied between the source and the drain electrodes. SPCM with a 1500 nm wavelength reveals signals near the source and drain electrodes. Photocurrent exhibits a strong dependence on the polarization of the incident light, which is explained by the linear dichroism of BP. We would like to note the large background photocurrent reported in the experiments. The background photocurrent is as large as 1 nA. Considering the signal itself and the fact that the authors employ the lock-in technique to only measure the photoinduced response, the background is very large, raising concerns regarding a pinhole leak to the oxidized Si substrate. To eliminate the possibility of the extrinsic anisotropic scattering of photo-excited carriers from the metal-electrode edge, the authors fabricated a ring-shaped metal electrode as a photocurrent collector. This would provide an isotropic collection of photogenerated carriers and eliminate the influence of the electrode edge.

Similar to the Hall-bar device, they observed a linear polarization-dependent photocurrent, with a large background photocurrent, coming from all over the scan area. The SPCM maps collected at different wavelengths show that the polarization dependence of the photocurrent emerges from the linear dichroism. The authors also discuss the origin of the photoresponse in detail. As we argued in previous paragraphs, the authors have considered photothermoelectric and photovoltaic effects in conjunction. The authors show that, near-zero source-drain bias, the photothermoelectric effect dominates the signal, while beyond ±150 mV, the photovoltaic response becomes more dominant. SPCM with 780 nm laser at 1.35 mW (~10 kW $cm^{-2}$) laser power is used for the measurements.

Na et al. [133] studied the tunneling transistor structure based on BP-$SnSe_2$ junction, Jiang et al. [134] studied the tunneling devices based on graphene-oxidized BP samples, and Wang et al. [135] studied BP-$In_2Se_3$ junctions using SPCM. However, we didn't discuss these studies as the SPCM measurements add little to the understanding developed by the earlier studies we discussed, but rather involve the device performance for the structures. To conclude, all these studies reveal how strong anisotropy in the materials structure can lead to an anisotropic photoresponse. Attention has to be paid to linear dichroism, that is, differences between the absorption of light polarized in different axes of the sample, while interpreting the SPCM results.

### 3.3. Photoresponse of metallic and semi-metallic 2D materials

As elaborated on previous parts, photoresponse from metallic and semi-metallic materials is possible via various photothermal mechanisms such as the Seebeck effect, Ettingshausen effect, and bolometric effect [136]. Thus, we find it instructive to discuss the studies on semi-metallic and metallic 2d materials.

We reported the first SPCM measurement on a metallic van der Waals material in a few-layer thick 3R-$NbS_2$ [114]. Our findings determine that the laser-induced heating, which is determined to be 40 mK/μW for a diffraction-limited 532 nm laser spot, is responsible for the observed photothermal effects. At zero bias, we observed a strong thermoelectric effect across the gold-$NbS_2$ junction, and at finite biases, the signal is dominated by the bolometric effect, thanks to the relatively high temperature coefficient of resistance of 3R-$NbS_2$ and poor thermal conductivity perpendicular to the van der Waals plane. The former ensures that there is a large enough local resistivity change to be measured, and the latter ensures that higher local temperatures are due to poor heat dissipation to the substrate. We also performed similar measurements on 2H-$TaS_2$ few-layered crystals and demonstrated very similar results [27,84]. The photoresponse in metallic two-dimensional materials is similar to what is observed in metallic NWs.

$Bi_2Se_3$ is a topological insulator, with a small band gap (~0.3 eV) and topologically protected conducting surface states. Thus, we would like to discuss SPCM studies on $Bi_2Se_3$ in this sub-section. We would like to pay particular attention to the study by Pournia et al. [137] where they have studied mechanically exfoliated ~100 nm thick crystals of $Bi_2Se_3$. They performed SPCM at high and low temperatures at 1070 nm laser with 150 μW (2.5 kW $cm^{-2}$) power. They show that the observed photoresponse can be explained by the photothermoelectric effect with ~30 mK temperature increase at the contact-$Bi_2Se_3$ junction. Moreover, they demonstrated that laser polarization modulates the measured photovoltage. This polarization dependence is a result of the boundary conditions for the electric field at the edge of a metal contact. The tangential component of the electric field is zero at the interface for even when the laser is located directly on the interface, while the perpendicular component has a more rapid fall off with slight strengthening of the field intensity. As a result, the parallel polarization results in larger heating of the metal-$Bi_2Se_3$ boundary. This is an important finding that should be considered when interpreting the SPCM results near metallic contacts and structures, in any material.

Another important SPCM study on $Bi_2Se_3$ is by Hou et al. [138] where they demonstrated millimeter-long transport of photogenerated carriers in Sb-doped nanoribbons. They performed SPCM at temperatures ranging from room temperature to 7 K. Notably, while the photocurrent decay length at 200 K is determined to be 3 μm, it increases to 300 μm at 7 K. Such long photocurrent decay is not observed in pure $Bi_2Se_3$ nanoribbons, as pure samples tend to be degenerately n-doped. Sb doping brings the Fermi level close to the Dirac point. This is attributed to the formation of superfluid-like exciton condensates, where carriers relax from the bulk to surface states to form scattering-free, coherent bosons. Similar to Cooper pairs in superconductors, these surface excitons open many-body energy gaps and propagate ballistically over long distances without generating an electrical current due to their charge neutrality. However, significant photocurrent is eventually produced when these excitons dissociate upon reaching the metal-TI contact. A pump-probe SPCM study on Sb-doped $Bi_2Se_3$ has also been recently reported by Silva et al. [11]. Their results are consistent with the formation of exciton condensates in topological insulators, but do not fully agree with the behavior expected from the hot carriers.

## 4. Systematic Interpretation of SPCM Results

When performing experiments with the SPCM, various parameters can be systematically controlled to probe the origin of the photoresponse. The list of external parameters that can be controlled is given in **Table 1**. These parameters are divided into four main categories: (1) light-, (2) electrode metal-, (3) substrate-, and (4) ambient-related. Each of these categories has several parameters that can be tuned for controlled experimentation. Some of these parameters are easier to tweak than others in an SPCM setup, and the effect of some parameters is more significant than others, as discussed in the previous sections. It is imperative to consider all the listed parameters in an SPCM experiment and interpret the results accordingly.

**Table 1**. A list of experimental parameters in SPCM measurements.

| **Light** | ***Parameter Range / Effect*** | **Substrate** | ***Parameter Range / Effect*** |
|---|---|---|---|
| Wavelength ($\lambda$) | *UV to Terahertz /Absorbance of the material, below or above the band gap of semiconductors* | Thermal conductivity | *Photothermal effects* |
| Power ($P$) | *CW or pulsed / various mechanisms can become dominant at various power levels* | Dielectric constant | *Dielectric variations can alter local electronics* |
| Polarization | *Linear, circular, or unpolarized* | Back-reflection | *Increased laser absorption* |
| Intensity modulation frequency ($f$) | *A few Hz to tens of kHz* | Thermal boundary conductance | *Modulation of photothermal effects* |
| | | Interfacial charges | *Local doping of materials* |
| **Contact Electrode** | | **Ambient** | |
| Work function | *Type of junction with the material: Ohmic, Schottky* | Pressure/Strain | *Can break symmetries and cause band shifts, alter the Seebeck coefficient* |
| Seebeck coefficient | *Thermoelectric response* | Temperature | *Modify interactions* |
| Bias voltage | *Control dominant effect* | Moisture | *Lead to surface-related effects, cooling* |
| Inductive and capacitive effects | | Oxygen | *Oxidation-related effects, charge transfer* |
| Contact boundary effects | *Position of the signal* | Gate voltage | *Controlling the electrostatic environment, charge density* |
| Contact resistance | *Affects the dominant photocurrent mechanism* | Magnetic field | *Controlling* |

Light excitation forms the basis of the SPCM technique, and tuning excitation-related parameters can provide important insight into the experimental results. Excitation light wavelength can be changed to tune the photon energy above or below the bandgap of a semiconducting sample. At different wavelengths, the absorbed energy may change along with the incident photon flux. Moreover, wavelength-dependent plasmonic resonances can be modulated for the metallic samples. However, at longer wavelengths, resolution will be lower due to the diffraction limit. Another very important light-related parameter is the power of the incident beam. The dependence of the photoresponse on the excitation power can provide an understanding of the origin of the photoresponse. However, the complex interplay of various effects typically complicates the interpretation of the results, as discussed in Au NW papers by the Natelson group [21,71]. Especially when there are temperature-dependent phenomena that can be affected by the laser-induced heating, non-linear intensity dependence is possible. It should be noted that in SPCM studies of low-dimensional materials, laser-induced heating can cause the local sample temperature to rise by tens of Kelvin or more. A proper interpretation of the laser power dependence should include a careful analysis of multiple photoresponse mechanisms.

Polarization of the excitation light is also an important parameter. Especially in materials where the inversion symmetry is broken due to the crystal or extrinsic plasmonic structures, photoresponse strongly depends on the incident polarization. As we have seen in the case of valley Hall effect, BP and $Bi_2Se_3$ studies, the contact-sample boundary and linear dichroism of the measured sample can lead to significant alterations in the observed SPCM. Thus, claims regarding the effect of polarization on the SPCM results should be supported by alternative measurements such as polarization-dependent absorption measurements. Moreover, substrates that would lead to increased photon absorption due to optical interference from thin film interfaces should be taken into account while interpreting the results.

Conventional SPCM requires electrical contacts to the samples, and these contacts may also play a role in the results. Especially when studying semiconducting samples, the nature of the contact, whether Schottky or Ohmic, determines the type of response near the material-contact boundaries. Moreover, contact resistance also plays a role in the observed response. For instance, large contact resistance typically suppresses bolometric signals and leads to large responses near the contacts. Large capacitive or inductive effects can also modify the measured signal amplitude due to the ac nature of the lock-in-based SPCM measurements. Despite many metals exhibiting low Seebeck coefficients, graphene/graphite electrodes or some metals can strongly influence the photoresponse measured by SPCM. Finally, contact geometry can cause suspended structures or plasmonic heating in the samples that may lead to enhanced response from the metal-semiconductor interface. Contact-free photocurrent measurement methodology has also been recently introduced [139,140]. Nitrogen-vacancy centers in diamond are used as proximal quantum magnetometers to spatially resolve and map the photogenerated current. Despite the technique being in its infancy, it can provide deeper mechanistic insight into photocurrent generation mechanisms.

Substrate is another important part of the SPCM-related studies. Often overlooked, substrate-sample interactions can significantly impact the SPCM measurements. For instance, trapped charges at the low-dimensional material-substrate interface can lead to local doping of the sample, which can lead to changes in the photoresponse. Moreover, phonon scattering can alter the Seebeck coefficient in different parts of the sample, leading to same-material thermoelectric junctions as we have discussed before. Dielectric constant variations or substrate screening effects are other parameters to be considered in SPCM measurements. Similarly, thermal-boundary conductance between the substrate and the sample, along with the thermal conductivity of the substrate, can significantly modify the contribution of the photothermal effects to the measured photoresponse.

As illustrated in many studies we discussed, ambient parameters are also important in interpreting SPCM results. Temperature, pressure/strain, and presence or absence of oxygen can strongly influence the photoresponse of a sample. These parameters can be altered in a controlled manner to elucidate the underlying response mechanism(s). Also, external electric and magnetic fields can be classified as ambient parameters since they are extrinsic to the sample. The effect of the electric field, especially in semiconductors and semimetals, is significant. Moreover, the photo-Nernst effect can be obtained via an external magnetic field.

All these parameters, whether controlled or not, have to be considered simultaneously while interpreting the SPCM measurement results. The fundamental reason is that, unlike bulk samples, the heat dissipation channels are limited in low-dimensional materials, and the local temperatures can rapidly increase. Moreover, a tightly focused laser beam in an SPCM measurement can lead to multiple effects in a material at the same time, with a similar response to most experimental parameters. For instance, as we have seen in $MoS_2$, gate modulation can simultaneously affect the photovoltaic and photothermal effects. Slight changes in the sample geometry, substrate type, and incident light polarization could also lead to changes in the light absorption of the samples. For instance, using a double-sided polished sapphire or a single-sided polished sapphire leads to slight changes in the SPCM results due to the difference in the back-reflection. To conclude, irrespective of the material type, a systematic sweeping of the parameters should be done before interpreting the SPCM results to avoid finding unrealistically long minority carrier lifetimes. Moreover, SPCM results should be complemented by a technique that doesn't involve light excitation for a better interpretation.

Another important aspect that requires attention in interpreting SPCM results is the theoretical and computational modelling of the experimental data. For 1D semiconducting structures, Fu et al. [141] provides a comprehensive theoretical modelling for SPCM studies. For a bipolar classical semiconductor, the total current conducted by electrons is given by the following expression [142,143]:

$$I_n(x) = -n(x)|e|\mu_n \left[ \frac{1}{e} \frac{dE_{Fn}(x)}{dx} + S_n(x) \frac{dT(x)}{dx} \right]$$

Here, $n(x)$ is the local electron density, $E_{Fn}(x)$ is the electron quasi-Fermi level, $\mu_n$ is the electron mobility, $S_n(x)$ is the local Seebeck coefficient and $T(x)$ is the local temperature. A similar equation can be written for the holes. Finally, a continuity equation for both electrons and holes describes the local current flow over the nanowire:

$$\mp \frac{1}{|e|} \frac{dI_{n,p}(x)}{dx} + R - G = 0$$

where negative and positive sign corresponds to electrons and holes, respectively. Laser illumination generated nonequilibrium charge carriers at a rate $G$ and they recombine at a rate $R$ via various mechanisms. A more comprehensive approach is provided in an earlier review paper by Graham et al. [29] both for 1D and 2D materials [144].

Depending on the type of contacts being Schottky, Ohmic, or Schottky on one side and Ohmic on the other side, photocurrent profiles show differences. This theoretical modelling and simulations reveal overlooked effects in scanning photocurrent microscopy and challenge standard analytical assumptions. Key findings indicate that Schottky-contact devices are effective for measuring minority carrier diffusion lengths, though using long nanowires or electrical bias is recommended for the best accuracy. Conversely, the internal built-in potential cannot be correctly determined by simply integrating photocurrent; doing so ignores the nonlocal nature of diffusion and leads to significant errors in estimating the electric field. Finally, the results disprove the assumption that Ohmic-Ohmic devices rely solely on diffusion transport. Instead, carrier mobility mismatches generate significant drift currents even without external bias, altering the expected voltage distribution [143].

We would like to note that the theoretical modelling provided above can be applied to classical semiconductors where non-equilibrium charge carriers can be described via single-electron models. In correlated materials where the energy gap emerges due to quasiparticle interactions, the provided model may fail [58,67]. Especially in cases where photothermal effects are dominant alongside Newtonian cooling, unrealistically long carrier lifetimes and minority diffusion lengths can be measured even in Si nanowires. This is particularly contrasted by EBIC and SPCM measurements. Implementation of the Shockley-Ramo theorem is also needed for comprehensive models with quantitative predictive capabilities [145]. Moreover response near the electrical contacts of gapless materials cannot be captured by the provided model [146].

## 5. Limitations of SPCM and Outlook

Despite the clear advantages SPCM offers in understanding light-matter interactions and photoresponse mechanisms in low-dimensional materials and their devices, SPCM has shortcomings as well. Conventional SPCM is based on the raster scanning of a laser beam focused down to a diffraction-limited spot. This sets a lower bound on the resolution of the photocurrent maps. This results in limited information below several hundred nanometers in diameter. Although SPCM with near-field optics is possible, the significant reduction in the signal intensity while coupling the beam through near-field optics limits its use. Another way around is to use tip-enhanced photocurrent microscopy [7,147]. For instance, in a study by Woessner et al. [7] tip-enhanced SPCM is used to study bare and encapsulated graphene flakes. With a resolution reaching down to 25 nm, the authors could map out the spatial distribution of photocurrent with exceptional detail. However, further studies are required to fully harness the capabilities of tip-enhanced SPCM.

Another limitation of SPCM is the limitations in discerning various photo-induced effects from each other. As we have discussed in previous sections, photoresponse contains signals from various effects, and separating contributions from these different effects often requires a careful theoretical analysis or other techniques to be used in conjunction with SPCM. For instance, photothermoelectric effects can be mixed with photovoltaic effects or non-local current generation. Especially in semiconducting low-dimensional materials, any experimental parameter within SPCM may not be able to resolve the source of the photoresponse, as various photocurrent-generating effects can get intermixed.

Despite the aforementioned limitations, there are multiple potential areas in which SPCM can be employed. As recently reviewed by Ma et al. [9], photocurrent generation is highly sensitive to physical processes across various spatiotemporal scales, making it an increasingly popular tool over the last ten years. It serves as a versatile method for analyzing quantum material properties, including electronic states, Bloch band geometry, and quantum kinetics. SPCM combined with other methods can be used as a probe of charge, spin, and collective excitations. As we have reviewed in the case of $Bi_2Se_3$ and graphene, SPCM can be used as a probe of quantum geometric photocurrents, and it is possible that it can find more use in exploring topological effects. Combining the measurement capabilities of nitrogen-vacancy centers in diamond with SPCM has also become popular in elucidating the photocurrent mechanisms in 2D materials in the past decade [139,148–150].

Another area where SPCM can be employed is thermal conductivity measurements in low-dimensional materials. As we have discussed in the previous sections, we used thermal conductivity as an input parameter to model the laser-induced heating in materials with bolometric or thermoelectric response [20,58,114]. However, recently, we realized that a reverse analysis can yield thermal conductivity as well [27,84]. This is particularly exciting as measuring the thermal conductivity of low-dimensional materials is challenging due to the limited thermometers available at the micro- and nanoscale. The method we introduced is based on the temperature-dependent resistivity change of the materials, namely the bolometric effect. We used SPCM to demonstrate that in partially suspended thin metallic sheets, local resistivity change can be used as a thermometer to extract the thermal conductivity via modeling. As the sensitivity in the SPCM is very high, even a small thermal gradient of less than a Kelvin can produce a large enough photoresponse to extract the thermal conductivity. This enables very sensitive measurement of thermal conductivity. We will not discuss the method in detail here as it is beyond the scope of this review. However, improvements to the method could provide an added advantage to SPCM as a versatile tool.

Although in this review, we focused mostly on single-material devices of 1D and 2D materials, heterostructures of these systems are very important for the discovery of new physics, alongside the applications [151,152]. As we have mentioned in various sections, the interface between low-dimensional materials, as well as the supporting substrate, plays a substantial role in photoresponse mechanisms, and the role of SPCM-based studies is going to increase in the following decade to have a better understanding of the optoelectronics of low-dimensional heterostructures. A recent paper by Xu et al. [115] demonstrates this point. They use a three-dimensional scanning near-field optical microscopy (SNOM) based SPCM with 4 nm spatial resolution. By varying the probe-to-sample distances, they could independently map photovoltaic and photothermoelectric contributions to the photoresponse of $MoS_2$. Moreover, they demonstrated that if a thin layer of h-BN is placed on the $MoS_2$ PTE response is enhanced. They attributed enhanced PTE response to the high in-plane thermal conductivity of h-BN that expands the hot spot at the interface substantially. There are multiple other studies related to the heterostructures of low-dimensional materials [153–156].

In summary, by reviewing the relevant literature, we establish that SPCM is a powerful technique for determining the optoelectrical response of one and two-dimensional materials. We also highlighted certain aspects in the interpretation of the SPCM measurements that should be taken into consideration. Depending on the electronic structure of the material or device under investigation, different properties such as minority carrier lifetime, responsivity, and thermal conductivity can be extracted using SPCM. This versatility can be combined with tip-enhanced techniques to achieve super-resolution. Moreover, methods like scanning thermal microscopy in conjunction with SPCM provide a superior experimental view into light-induced photoresponse in low-dimensional materials. SPCM can provide time-resolved information regarding the photoresponse by incorporating ultra-fast time-resolved methods. The ability to measure thermal conductivity of metallic and correlated materials also exhibits its versatility as a measurement tool. Overall, SPCM is a versatile tool with many potential improvements and applications on various materials and devices

## 6. Acknowledgements

T.S.K. acknowledges support from TUBITAK under grant no 123F129.

## 7. Appendix A – Experimental Implementation of SPCM

### 7.1. The electrical configuration

Two or four electrodes to the material are required for measuring the electrical photoresponse. As we further elaborated later, depending on the mechanisms involved in the photoresponse and the material properties, photocurrent can emerge over a very large range, from a few picoamperes to milliamperes. Equivalently, photovoltage can be as low as a few nanovolts up to a few millivolts. This requires the SPCM to have a large measurement range to prevent saturation of the measured signal. It is also important to eliminate the effects of extrinsic experimental conditions on measured signals, which can lead to experimental artifacts that are not correlated with the incident optical excitation. Moreover, especially in cases where voltage biasing is experimentally required, the dark current can be multiple orders of magnitude larger than the photocurrent. Due to the aforementioned reasons, as well as to improve the signal-to-noise ratio, typically lock-in amplifiers (LIA) are used in the SPCM. The excitation laser intensity is modulated at a fixed frequency via mechanical or electrical methods, and the modulation frequency is fed to the LIA for frequency-dependent signal amplification. LIA measurements also ensure that only the photo-induced electrical changes are measured, with a very large dynamic range, even when the signal is three orders of magnitude smaller than the noise, while the DC sources are omitted. However, there are low cost implementations of SPCM without the use of LIAs reported in the literature [157].

A current preamplifier could be employed to filter and amplify the signal before the LIA at the cost of reducing the measurement bandwidth. Using a current pre-amplifier also ensures sufficiently high input impedance to the LIA, and the bias voltage or current can be applied through the pre-amplifier. However, the chopping frequency and the pre-amplifier filter settings must be set carefully to avoid attenuating the measured signal. Signals down to a few femtoamperes can be measured with a proper choice of equipment, shielding of the measurement area, and appropriate cabling (i.e., cables fixed to reduce triboelectric effects, triaxial shielding, etc.). A second LIA is required to measure the signal from the photodiode (PD) and create the reflection map. Without LIA, PD will register intensity modulation due to the chopped laser. For PD, a software-based phase-locked loop can also be used to reduce the equipment cost, as PDs are typically sensitive and have their built-in amplification circuitry. **Figure 1a** shows the electrical configuration of a typical SPCM setup.

### 2.1.1. Laser beam steering methods

Raster or meander scanning of a focused laser spot is the essence of SPCM measurements. In a typical setup, laser spot scanning is performed either by the sample stage or by the galvo mirrors. The scan area is pixelated, and to achieve the desired spatial resolution at the diffraction limit, the steps between each pixel should be less than ~250 nm. Also, the accuracy of positioning the laser spot should be better than approximately half of the wavelength of the laser to provide scanning resolution at the diffraction limit. Closed-loop piezoelectric scanners can be used to achieve the required resolution and accuracy. In both stage and galvo scanning, the total scan area is limited to a few tens of micrometers. There are also less common scanning strategies, such as scanning the focusing objective with respect to the incident beam or the sample. Different scanning strategies are depicted in **Figure 13**.

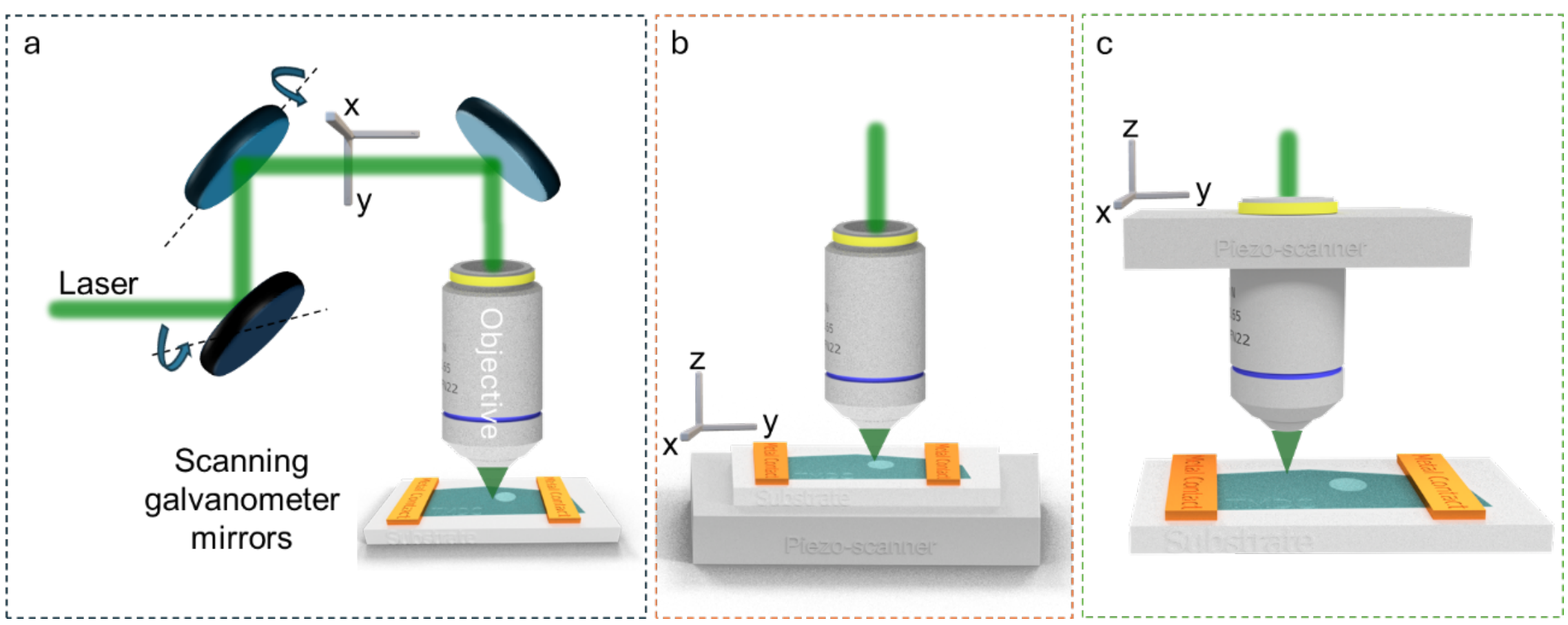

**Figure 13 a.** Scanning galvanometer mirrors steer the laser beam over the microscope objective in the desired pattern. As a result, the focused light spot scans the sample. **b.** Piezoelectric scanners scan the sample while the objective and the rest of the optics are fixed. Allows multiple beam excitation. **c.** The piezoelectric scanner scans the objective over the sample with respect to the incident laser beam. As a result, the focused spot is scanned over the sample.

The physical position of the laser spot must be coordinated with the recorded photocurrent data. It is also imperative to record and exhibit photoresponse and the reflected light intensity data in real-time by a computer or a plotter. The real-time plotting of the data is important as certain adjustments on the LIA parameters can be made to prevent saturation of the output or the input signal and to monitor the phase of the measured signal. Typically, custom software and hardware solutions are required for experimental instrument-computer interfacing and data recording.